\documentclass[traditabstract]{aa} % for the abstract without structuration
\usepackage{graphicx}
\usepackage{txfonts}
\usepackage{lscape}
\usepackage{float}
\newcommand{\mic}{$\mu$m}
\newcommand{\ang}{$\rm \AA$}

\newcommand{\lbol}{$L_{\mathrm{bol}}$}

\newcommand{\Lsun}{$L_{\odot}$}
\newcommand{\Msun}{M$_{\odot}$}

\newcommand{\ra}[0]{$\rightarrow$}

\newcommand{\hh}[0]{H$_2$}

\newcommand{\Oone}{[O\,{\sc i}]}
\newcommand{\Odrie}{[O\,{\sc i}]\,63}

\newcommand{\tgas}{$T_{\mathrm{gas}}$}
\newcommand{\tdust}{$T_{\mathrm{dust}}$}
\newcommand{\trot}{\mbox{$T_{\mathrm{rot}}$}}
\newcommand{\tvib}{$T_{\mathrm{vib}}$}

\newcommand{\luv}{$L_{\mathrm{UV}}$}
\newcommand{\lpah}{$L_{\mathrm{PAH}}$}
\newcommand{\lacc}{$L_{\mathrm{acc}}$}
\newcommand{\eup}{$E_{\mathrm{up}}$}
\newcommand{\jup}{$J_{\mathrm{up}}$}
\newcommand{\mean}[1]{\mbox{$\langle#1\rangle$}}

\begin{document}

   \title{DIGIT survey of far-infrared lines from protoplanetary discs: II. CO
    \thanks{  {\it Herschel} is an ESA space observatory with science instruments
provided by European-led Principal Investigator consortia and with important
participation from NASA.}}

   %\subtitle{An overview}

   \author{Gwendolyn Meeus\inst{1}, Colette Salyk\inst{2}, Simon Bruderer\inst{3,4},  Davide Fedele\inst{3}, 
   Koen Maaskant\inst{4}, Neal J. Evans II\inst{5}, Ewine F. van Dishoeck\inst{3,4}, Benjamin Montesinos\inst{6},
   Greg Herczeg\inst{7}, Jeroen Bouwman\inst{8}, Joel D. Green\inst{5},  Carsten Dominik\inst{9},  
   Thomas Henning\inst{8}, Silvia Vicente\inst{10}, 
          \and
         the DIGIT team
          }

   \institute{Universidad Autonoma de Madrid, Dpt. Fisica Teorica, Campus
Cantoblanco, Spain \email{gwendolyn.meeus@uam.es}
           \and
           National Optical Astronomy Observatory, 950 N. Cherry Avenue,
     Tucson, AZ 85719, USA  %2
           \and
           Max Planck Institute f\"ur Extraterrestriche Physik,
           Giessenbachstrasse 1, 85748 Garching, Germany  %3
           \and
           Leiden Observatory, Leiden University, P.O. Box 9513, 2300 RA Leiden,
          The Netherlands %4
           \and
           The University of Texas at Austin, Department of Astronomy,
2515 Speedway, Stop C1400, Austin, TX 78712-1205, USA  %5
            \and
            Dept. of Astrophysics, CAB (CSIC-INTA), ESAC Campus, P.O. Box 78,
28691 Villanueva de la Ca\~nada, Spain %6 BM
           \and
            Kavli Institute for Astronomy and Astrophysics, Yi He Yuan Lu 5,
             Beijing, 100871, P.R. China, %7
           \and 
           Max Planck Institute for Astronomy, K\"onigstuhl 17, 69117,
           Heidelberg, Germany  %8
           \and
           Anton Pannekoek, Amsterdam %9
          \and
           Kapteyn Astronomical Institute, Postbus 800, 9700 AV Groningen,
            The Netherlands %10
             }

   \date{Received ; accepted }

%++++++++++++++++++++++++++++++++++++++++++++++++++++++++++++++++++++++
%++++++++++++++++++++++++++++++++++++++++++++++++++++++++++++++++++++++
% \abstract{}{}{}{}{}
% 5 {} token are mandatory
\authorrunning{Meeus et al.}
\offprints{Gwendolyn Meeus,\\ \email{gwendolyn.meeus@uam.es}}

\abstract{CO is an important component of a protoplanetary disc as it is one of the most abundant
gas phase species. Furthermore, observations of the CO transitions can be used as a diagnostic
of the gas, tracing conditions in both the inner and outer disc. We present {\em Herschel}/PACS spectroscopy of a
sample of 22 Herbig Ae/Be (HAEBEs) and 8 T Tauri stars (TTS), covering the pure rotational CO transitions
from $J$ = 14 \ra \ 13 up to $J$ = 49 \ra \ 48. CO is detected in only 5 HAEBEs, AB Aur, HD 36112,
HD 97048, HD 100546 and IRS 48, and in 4 TTS, AS 205, S CrA, RU Lup, and DG Tau. The 
highest transition seen is $J$ = 36 \ra \ 35, with \eup \ of 3669 K, detected in HD 100546 and DG Tau. We 
construct rotational diagrams for the discs with at least 3 CO detections to derive \trot, and find average 
temperatures of 270 K for the HAEBEs and 485 K for the TTS. HD 100546 requires an extra temperature 
component with \trot \  $\sim$ 900-1000 K, suggesting a range of temperatures in its disc atmosphere,
consistent with thermo-chemical disc models. 
In HAEBEs, the objects with CO detections all have flared discs in which the gas and dust are thermally 
decoupled. We use a small model grid to analyse our observations and find that an increased amount of 
flaring means higher line flux, as it increases the mass in warm gas. CO is not detected in our flat discs
as the emission is below the detection limit. We find that HAEBE sources with CO detections have high \luv \ 
and strong PAH emission, again connected to the heating of the gas. In TTS, the objects with CO
detections are all sources with evidence for a disc wind or outflow. For both groups of objects,
sources with CO detections generally have high UV luminosity (either stellar in HAEBEs or due to accretion 
in TTS), but this is not a sufficient condition for the detection of the far-IR CO lines. 
}

  \titlerunning{CO in protoplanetary discs}
  \authorrunning{G. Meeus et al.}

   \maketitle

%++++++++++++++++++++++++++++++++++++++++++++++++++++++++++++++++++++++
%++++++++++++++++++++++++++++++++++++++++++++++++++++++++++++++++++++++
\section{Introduction}
\label{s_intro}

It is firmly established that circumstellar discs around young stars are sites of planet formation.
During the first 10 Myr, the initially gas-rich disc will first evolve into a transitional and finally into
a debris disc, while dispersing its gas content. The knowledge of this dispersal process and what
favours/hinders it is a crucial part of the planet-formation puzzle, as the amount of gas present in a
disc determines whether gas-rich planets can form. Therefore, three aspects 
need to be characterised well: the disc geometry, the solid state content, and the gas content. Both 
the geometry and the dust in HAEBE discs are well-studied (e.g. the review
by Williams \& Cieza \cite{williams2011}; Meeus et al. \cite{meeus2001}, Dominik et al. \cite{dominik2003} 
and Benisty et al. \cite{benisty2010} for the geometry; Acke et al. \cite{acke2010} for the Polycyclic Aromatic 
Hydrocarbon (PAH) and Juh\'asz et al. (\cite{juhasz2010}) for the dust properties).

To understand the gas component --- spatial distribution, chemistry and physical properties --- 
it is necessary to observe a range of transitions in several species, as they can originate in different 
regions of the disc under distinct conditions (density, temperature, radiation field). For the purpose 
of studying the gas disc, \hh \ and CO lines are most often used, since they are the most abundant 
molecules present. In the FUV, \hh \ probes the warm gas and is frequently detected in TTS (e.g., 
Herczeg et al. \cite{herczeg2006}; 
Ingleby et al. \cite{ingleby2011}; France et al. \cite{france2012}). In the IR, the detection of \hh \ is more
difficult, due to the nature of the \hh \ molecule (homonuclear with wide spacing of energy levels). In a 
survey of 15 HAEBEs with CRIRES, Carmona et al.\  (\cite{carmona2011}) detected ro-vibrational 
transitions at 2.1218 \mic \ in only two objects. Earlier, Bitner et al.\ (\cite{bitner2008}), Carmona et al.\ 
(\cite{carmona2008}), Martin-Za\"idi et al.\ (\cite{claire2009,claire2010}) searched for pure rotational 
lines of \hh \ at 17.035~\mic \ in a sample of 20 HAEBEs; only two detections were reported. In TTS,
the success rate is a little higher: Lahuis et al. (\cite{lahuis2007}) detected mid-IR \hh \ emission 
lines in six TTS, while Bary et al. (\cite{bary2008}) found five more \hh \ detections in the near-IR.

In sharp contrast, the detection of CO, although less abundant, is much easier. Its lines have been used
to trace both the inner and outer disc. 
The fundamental ro-vibrational CO band ($\Delta$v$ = 1$) at 4.7 \mic \ traces warm ($T >$ 100 K) gas in
the terrestrial planet-forming region (0.1 -- 10 AU). This band is often observed in HAEBEs (e.g. Blake \& Boogert
\cite{blake2004}; Brittain et al. \cite{brittain2007}) and in TTS (e.g. Najita et al. \cite{najita2003}; Salyk et al.
\cite{salyk2011}). 
The bands are rotationally excited up to high $J \,  ( >$ 30), resulting in \trot \ between 900 and 2500 K.
\trot \ is much lower for $^{13}$CO, $\sim$ 250\,K (Brown et al. \cite{brown2013}).
UV fluorescence can cause super-thermal level populations, as observed (Brittain et al. \cite{brittain2007}) and
modelled (Thi et al. \cite{thi2013}) in UV-bright HAEBEs where \tvib $>$ 5000 K, as well as in TTs (Bast et al. 
2011; Brown et al. \cite{brown2013}). 
While some HAEBEs have CO extending to the dust sublimation radius (Salyk et al. \cite{salyk2011}), others 
have evidence for significant inner disc gas clearing (Goto et al. \cite{goto2006}, Brittain et al. \cite{brittain2007}, 
Pontoppidan et al. \cite{pontoppidan2008}, van der Plas et al. \cite{plas2009}).  
Thi et al. \cite{thi2013}, and van der Plas (Ph.D. Thesis) find that, in flaring discs, \trot \ $<$ \tvib, and the inner radius 
of CO, $r_{in}\sim$ 10 AU, while in self-shadowed discs, \tvib \ $\lessapprox$ \trot \ and $r_{in} \sim$ 1 AU: flaring 
HAEBE discs thus have lower CO abundances in their inner regions than flat discs. 
In a recent paper, Maaskant et al. (2013) show that several group I sources (the Õflaring discsÕ) have 
dust-depleted gaps separating an optically thin inner disc and a flaring outer disc.  It has been proposed 
that dust-depleted regions are a typical property of group I discs. This geometry allows more radiation reaching 
the outer disc, so they can be heated more when compared to the group II (Õflat discsÕ). For the TTS, Brown et 
al. (\cite{brown2013}) found that most of the line profiles in TTS indicate that the 
CO emission originates in the disc plus a slow disc wind.

On the other end of the spectrum, CO low-$J$ pure rotational transitions ($\Delta$v$= 0$) can be observed in the 
millimeter. These low-$J$ transitions originate from cold, optically thick CO located in the outer disc, and are 
routinely detected in HAEBE (e.g. Thi et al. \cite{thi2001}, Dent et al. \cite{dent2005}) and in TTS (e.g. 
Koerner \& Sargent \cite{koerner1995}; \"Oberg et al. \cite{oberg2010}). As the lines are optically thick, the 
outer disc radius can be estimated from the line profile (e.g. Dent et al. \cite{dent2005}, Pani{\'c} et  al.
\cite{panic2008}).  In HAEBE discs, the dust temperature in the entire disc is high enough (at least
30 K) to avoid freeze-out of CO on the dust grains (e.g. Panic et al. \cite{panic2009}, Bruderer et al.
\cite{bruderer2012}), while in TTS, this happens more frequently (e.g. Hersant et al. \cite{hersant2009}).

The much less observed far-IR wavelength range covers rotationally-excited mid-to high-$J$ CO lines
which probe temperatures intermediate between the low-$J$ sub-mm and the vibrationally-excited near-IR lines.
Giannini et al. (\cite{giannini1999}) analysed ISO/LWS observations of 3 HAEBEs, covering rotational
lines from \jup = 14 to \jup = 19. Their modelling showed that the CO emission originates from a compact
region (up to a few 100 AU) around the central star, and suggest that the stellar FUV radiation excites the CO. 

CO far-IR lines  in HAEBEs and TTS were also observed by DIGIT (Sturm et al. \cite{sturm2010}) and 
GASPS (Meeus et al. \cite{meeus2012}) with the instrument PACS (Poglitsch et al. \cite{poglitsch2010}) 
onboard {\em Herschel} (Pilbratt et al. \cite{pilbratt2010}). Bruderer et al. (\cite{bruderer2012}; hereafter BR12) 
modelled pure rotational CO transitions in HD 100546 (Sturm et al. \cite{sturm2010}). They showed that, in this disc, 
1) in the disc surface, the gas temperature is decoupled from the dust temperature; 2) the low-$J$ ($J <$ 10) lines 
trace a radius between 70-220 AU, at a low scale height in the disc, the mid-$J$ ($J <$ 20) lines trace a radius of 
35-80 AU, while the high-$J$ ($J >$ 20) lines trace the upper disc atmosphere at 20-50 AU. 
Therefore, the higher the $J$, the closer to the star is the region the transition traces; 3) the highest $J$ lines are 
optically thin and scale with the amount of hot gas in the disc surface, while the lowest $J$ lines are optically thick 
and scale with the outer disc radius. The authors also show that the rotational diagram of HD 100546 (including, 
besides PACS also APEX data of low-$J$ transitions) can be fitted with 3 components: \trot $\sim$ 60, 300 and 750 K. 

We now present PACS spectra of a much larger sample of HAEBEs and TTS observed within the DIGIT Open Time 
Key Project (OTKP; P.I. Neal Evans). These spectra show pure rotational transitions of CO with $J_{\rm{up}}$ from 
14 up to 49, corresponding to \eup \ between 581 and 6724 K. Detections of O\,{\sc i}, C\,{\sc ii}, OH, H$_2$O, 
CH$^+$ are reported in a companion paper by Fedele et al. (submitted; paper I). In Sect.~\ref{s_sampobs}, we 
present the sample and the observations, while in Sect.~\ref{s_res} we show the line detections and  fluxes 
and construct rotational diagrams. In Sect.~\ref{s_ana} we look into the UV luminosity and  use a small disc model 
grid to explain our observations. In Sect.~\ref{s_disc}, we speculate on the impact of PAH luminosity and inner disc 
clearing on the CO emission, while in Sect.~\ref{s_conc}, we present our conclusions.

%++++++++++++++++++++++++++++++++++++++++++++++++++++++++++++++++++++++
\section{Sample and observations}
\label{s_sampobs}

Our sample consists of 22 Herbig Ae/Be stars and 8 T Tauri stars. The HAEBE sample has spectral types 
between B9 and F4, and is representative for the HAe stars, while the TTS sources tend to be distinct from typical 
TTS in some way (such as having a large amount of veiling, or the presence of a strong disc wind or outflow or a 
disc gap). We refer to Paper I for a discussion of the sample properties.

For each HAEBE star, we derived its disc group (I: flared discs, II: flat discs, in the classification of Meeus 
et al. \cite{meeus2001}), and calculated \luv \ by integrating the UV flux obtained from the IUE archive 
between 1150 and 2430 \ang \ for the HAEBEs. For the TTS, we measured \luv \ between 1250-1700 \ang \
directly for a few sources. 
When direct measurements of FUV were not available, they were derived instead from observations 
of the \ion{H}{I} Pf$\beta$ line and an empirical relationship between \lacc \ and Pf$\beta$ (Salyk et 
al. \cite{salyk2013}), in conjunction with an empirical relationship between \lacc \ and \luv \ (Yang et al.  
\cite{yang2012}). These results are listed in Table~\ref{t_para}.

\begin{table}
\caption{Some properties of the sample. }
\begin{center}
\begin{tabular}{lcrrr}
\hline\hline
HAEBE                  & Disc    & \luv / \Lsun                 &\lpah         & CO \\
                                & group & 1150-2430 \ang     &(\Lsun)        &det.\\
\hline
AB Aur                   & I           & 4.63                             &0.203 & Y\\
HD 35187             & II          & 2.23                             &0.034 & n\\
HD 36112             & I           & 1.32                             &0.029 & Y\\
HD 38120             & I           & --                                  &0.116 & n\\
HD 50138             & II          & --                                  &--                      & n\\
HD 97048             & I           & 7.69                             &0.367 & Y\\
HD 98922             & II          & --                                  &--                     & n\\
HD 100453           & I           & 0.29                             &0.038& n\\
HD 100546           & I           & 7.22                             &0.098 & Y\\
HD 104237           & II          & 1.54                             &--                     & n\\
HD 135344 B       & I           & $>$ 0.11$^{\rm a}$   &0.015 & n\\
HD 139614           & I           & 0.39                             &0.022 & n\\
HD 141569 A       & II/TO    & 6.83                             &0.007& n\\
HD 142527           & I           & $>$ 0.15$^{\rm a}$   &0.149 & n\\
HD 142666           & II          & 0.37--0.68                   &0.028& n\\
HD 144432           & II          & --                                   &0.003& n\\
HD 144668           & II          & 1.55--2.94                   &--                      & n\\
IRS 48                    & I           & 4.63                             &0.386& Y\\ 
HD 150193           & II          & 8.53                             &--                     & n\\
HD 163296           & II          & 3.21--5.58                  &--                      & n\\
HD 169142           & I           & 0.45                             &0.093  & n\\
HD 179218           & I           & --                                  &--                     & n\\
\hline
\hline
TTS                        &log \lacc /\Lsun                                & \luv / \Lsun     &CO  \\
                                &                                                           &1250-1700 \ang&det.\\
\hline
DG Tau                 &-0.13                                                  & 0.016$^b$        & Y\\
HT Lup                  &-0.65                                                  & 0.006$^b$        & n\\
RU Lup                 &-0.42                                                  & 0.01                   & Y\\
RY Lup                 &--                                                         & 0.002                 & n\\
AS 205 A/B          &-0.15/-0.37                                        & 0.016/0.01        & Y\\
SR 21                   &--                                                          & 0.0003$^b$     & n\\
RNO90                 &-0.16                                                   & 0.016$^b$       & n\\
S CrA A/B             &-0.10/-0.59                                        & 0.016/0.006$^b$ & Y\\
\hline
\end{tabular}
\end{center}
\label{t_para}
\tablefoot{HAEBE disc group classification from Meeus et al. (\cite{meeus2001}), Acke et al. (\cite{acke2010}) and 
Maaskant et al. (\cite{maaskant2013}). 'TO' stands for 'Transition Object', indicating that it lacks a near-IR excess. 
\luv \ is determined differently for the HAEBEs and the TTS, as described in the text (see Sect.~\ref{s_sampobs}); 
\tablefoottext{a}{\luv \  measured on the model photosphere (no UV observations available);}  \tablefoottext{b}{\luv \ 
estimated from accretion luminosities. \lpah \ is from the survey by Acke et al. (\cite{acke2010}) which includes our 
HAEBE sample; with '--' we indicate a non-detection. In the column 'CO det.' we indicate whether there was CO 
emission detected in our PACS spectra (Y) or not (n).}}
\end{table}

\begin{figure*}
   \resizebox{\hsize}{!}{\includegraphics {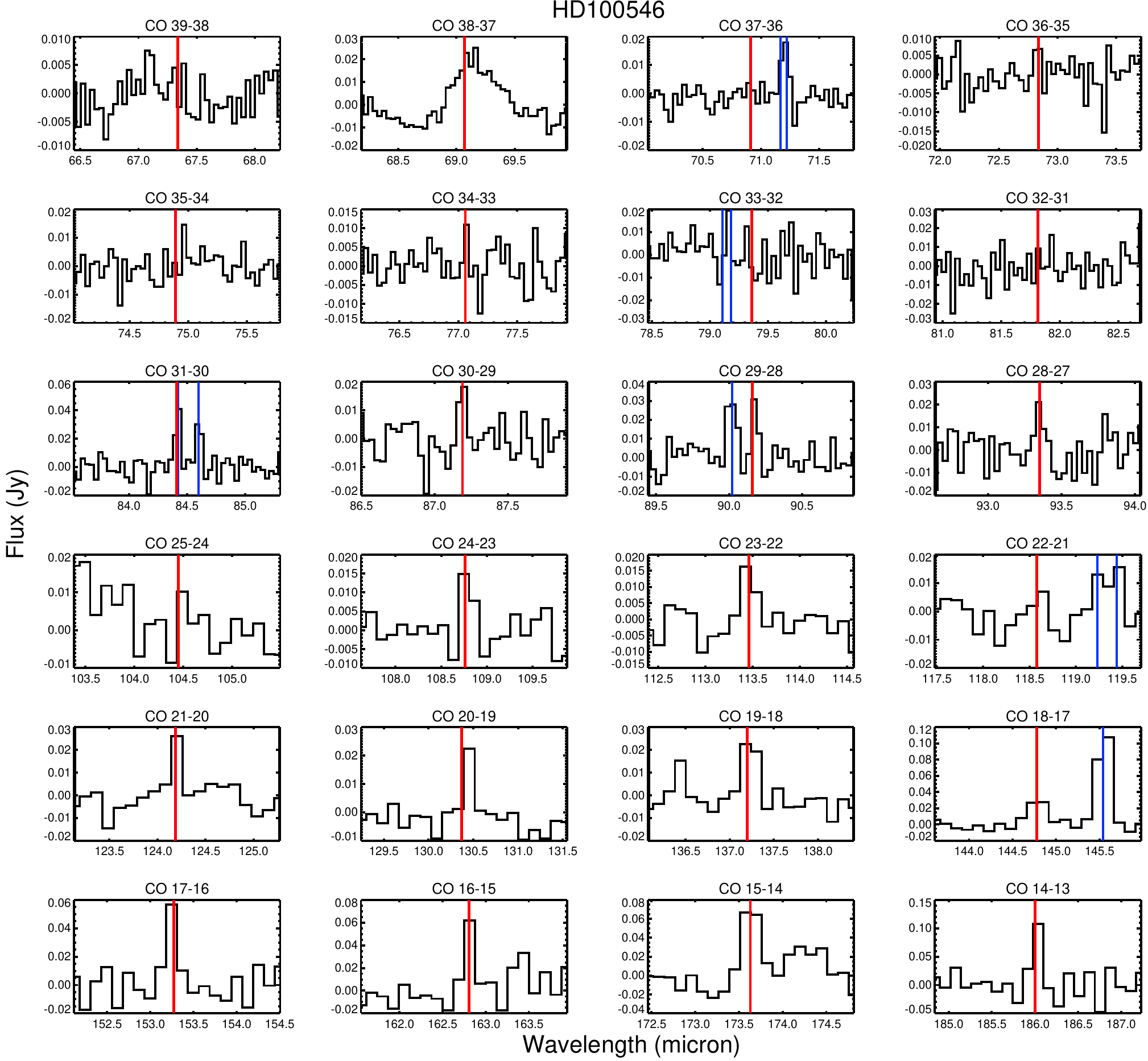}}
\caption{The spectra of HD 100546, centered on the CO lines, indicated in red. The OH line at 84.42 \mic \ blends with the 
CO $J$ = 31 \ra \ 30 line. In blue, we indicate the positions of other lines (CH$^{+}$ at 90 \mic \ and \Oone \ at 145 \mic; rest 
are OH doublets).
}
\label{f_spectra}
\end{figure*}

We obtained {\em Herschel}/PACS spectroscopy in SED range mode, covering the spectral range
between 51 and 210 \mic. Not all sources were observed in the full range. For an overview of the
observational settings and the reduction methods we refer to Paper I, where we also show the full
range spectra for each object.  In Fig.~\ref{f_spectra} we present a zoom in at the positions 
of the CO lines for HD 100546, while we show similar plots in Figs.~\ref{f_spectra1} to \ref{f_spectra8}
for those objects where CO was detected.

With the sensitivity of our spectra, we are able to achieve a typical S/N on the continuum of 100 and to detect a line 
with 3 $\sigma$ confidence when the line flux is higher than 1$\times$ 10$^{-17}$ W\,m$^{-2}$, varying 
with the observed wavelength and level of continuum flux. Furthermore, at the edges of the bands, the spectra 
become more noisy, resulting in higher upper limits (for a more detailed discussion on this topic we refer to the DIGIT 
data overview paper by Green et al. \cite{green2013}). For detected lines, we extracted the line fluxes using a Gaussian 
fit to the emission lines,
using the RMS on the continuum (excluding the line) to derive a 1$\sigma$ error on the line flux.
For non-detections, we give a 3$\sigma$ upper limit, also calculated from the continuum RMS. The derived line fluxes 
are listed in Table~\ref{t_lineflux}. The second row for each source is the flux density in the adjacent continuum.

\begin{table*}
\caption{Observed CO line fluxes in W\,m$^{-2} \times 10^{-18}$ (first row) and adjacent continuum flux density in Jy (second row). }
\begin{center}
\begin{tabular}{lcccccccccc}
\hline\hline
Transition  & $J$=38 \ra \ 37  &$J$=37 \ra \ 36   &$J$=36 \ra \ 35  &$J$=35 \ra \ 34   &$J$=34 \ra \ 33  &$J$=33 \ra \ 32
 &$J$=32 \ra \ 31   &$J$=31 \ra \ 30$^a$ \\
\hline
AB Aur        &$<$ 100.5  &$<$ 95.3    &$<$ 100.7  &$<$ 73.8     &$<$100.3 &$<$
79.5   &$<$ 77.9   &92.1 (2.6)   \\
                     &134.8 Jy     &134.3 Jy    &135.5 Jy     &136.6 Jy
&134.8 Jy   &133.1 Jy    &129.7 Jy    &132.0 Jy     \\
HD 36112  &$<$ 21.8    &$<$ 25.2   &$<$ 38.4      &$<$ 24.3  &$<$ 20.3   &$<$
25.1    &$<$ 24.9   &$<$ 29.5 \\
                     &20.5 Jy       &20.5 Jy      &20.6 Jy         &20.3 Jy
 &20.3 Jy      &20.0 Jy      &19.1 Jy       &19.1 Jy    \\
HD 97048  &$<$ 42.8    &$<$ 45.7   &$<$ 38.2     &$<$ 33.1   &$<$ 42.4   &$<$
37.6    &$<$ 40.2   &56.6 (15.0) \\
                     &65.8 Jy       &66.6 Jy      &68.1 Jy        &68.5 Jy
&70.3Jy      &69.1 Jy       &66.0 Jy      &68.5 Jy    \\
HD 100546&$<$ 107.4  &$<$ 82.9    &52.9 (12.6) &$<$ 44.1   &53.1 (14.4)&$<$
84.5  &$<$ 65.3   &139.7 (27.6)\\
                      &172.0 Jy    & 162.4 Jy   &160.6 Jy      &159.5 Jy
&156.7 Jy     &148.6 Jy   &139.7 Jy   &141.7 Jy   \\
IRS 48        &$<$ 28.7    &$<$ 23.0   &$<$ 37.5     &$<$ 31.2   &$<$ 28.4
&$<$ 37.9    &$<$ 38.6 &$<$ 33.7\\
                    &48.9 Jy       &47.2 Jy      &46.4 Jy        &46.6 Jy
&45.1 Jy       &42.6 Jy       &39.8 Jy    &39.4 Jy\\
AS 205        &$<$ 22.8    &$<$ 33.1   &$<$ 37.0     &$<$ 29.8   &$<$ 24.1
&$<$37.2    &$<$ 25.9 &46.0\\
                     &20.7 Jy       &20.5 Jy      &21.0  Jy       &20.8 Jy
&20.9  Jy      &20.7 Jy      &19.3 Jy     &20.4 Jy\\
S CrA          &$<$ 24.0    &$<$ 33.2   &$<$ 37.9     &$<$ 28.3   &$<$ 30.4
&$<$ 42.6   &$<$ 30.8  &$<$ 40.8\\
                     &23.6 Jy      &24.0 Jy       &24.5 Jy       &24.3 Jy
&24.6 Jy       &24.4 Jy      &23.7 Jy    &24.5\\
DG Tau      &$<$ 15.4    &$<$ 42.4   &25.7 (6.6)   &$<$ 18.9   &24.7 (5.5)
&$<$ 47.6     &19.8: (7.0)&128.3 (25.3)\\
                    &26.5 Jy       &26.6 Jy      &26.7 Jy        &26.6 Jy
&26.7 Jy       &26.1 Jy       &25.2 Jy     &26.2 Jy\\
\hline
\hline
                     &$J$=30 \ra \ 29  &$J$=29 \ra \ 28  &$J$=28 \ra \ 27       &$J$=25 \ra \ 24
&$J$=24 \ra \ 23   &$J$=23 \ra \ 22     &$J$=22 \ra \ 21  \\
\hline
AB Aur        &$<$ 57.8     &$<$ 81.5   &$<$ 55.5       &$<$ 54.2     &$<$ 47.0
   &58.9 (12.4)  &$<$ 45.7     \\
                     &133.0 Jy     &129.8 Jy    &123.6Jy         &103.7 Jy
&99.2 Jy      &93.8 Jy         &88.0 Jy      \\
HD 36112  &$<$27.6     &$<$ 22.7    &$<$ 32.3       &$<$ 17.7    &$<$ 13.3
&$<$ 11.0      &$<$ 9.1   \\
                     &19.7 Jy       &19.4  Jy      &19.0 Jy          &15.9 Jy
    &15.5 Jy       &14.6 Jy         &13.7  Jy   \\
HD 97048  &$<$ 43.0    &$<$ 32.5    &$<$ 42.7       &$<$ 36.4    &$<$ 30.4
&30.1  (8.7)   &14.7: (6.6) \\
                     &71.1 Jy       &70.7 Jy       &68.2 Jy          &61.5 Jy
    &59.1 Jy        &57.4 Jy         &53.7 Jy    \\
HD 100546&73.4 (14.9)&78.6 (18.7)&81.5 (11.2)    &$<$77.1     &71.3 (11.5)
&78.4 (11.2)  &$<$ 42.0  \\
                     &139.6 Jy     &135.6 Jy     &126.1 Jy       &107.9 Jy
&100.5 Jy      &91.5 Jy         &83.2 Jy     \\
IRS 48        &$<$ 26.0    &$<$ 29.0    &$<$  27.3      &$<$  22.7  &$<$  17.8
   &$<$ 13.0       &$<$  14.2\\
                    &39.0 Jy       &36.8 Jy       &34.4 Jy          &28.3 Jy
  &25.7 Jy         &23.2 Jy          &20.8 Jy\\
AS 205       &$<$ 28.3     &$<$ 20.8   &55.1  (7.3)     &18.9 (5.4)  &15.2
(4.8)    &30.1 (3.7)     &17.9 (3.8)\\
                     &19.8 Jy       &20.1            &19.1 Jy          &15.9 Jy
     &15.4  Jy       &14.5 Jy          &13.7\\
S CrA          &$<$ 24.6    &$<$ 23.2    &24.9 (6.4)      &26.6 (6.5)  &23.1
(5.7)   &42.2 (4.9)      &31.4 (10.6)\\
                    &24.4 Jy        &24.3 Jy       &23.6 Jy          &21.2 Jy
   &20.3 Jy        &19.0 Jy          &17.6 Jy\\
DG Tau      &36.4 (6.2)  &41.3 (7.3)   &51.5 (6.9)      &39.8 (4.8)  &38.4
(4.1)    &64.9 (3.3)      &40.2 (8.4)\\
                    &26.5 Jy      &26.0 Jy        &25.5 Jy          &21.4 Jy
   &20.5 Jy        &19.4 Jy          &18.5 Jy\\
\hline
\hline
                     &$J$=21 \ra \ 20   &$J$=20 \ra \ 19    &$J$=19 \ra \ 18    &$J$=18 \ra \ 17
&$J$=17 \ra \ 16   &$J$=16 \ra \ 15    &$J$=15 \ra \ 14   &$J$=14 \ra \ 13  \\
\hline
AB Aur        &$<$ 35.0  &39.2 (9.3)   &$<$ 29.8      &49.6 (10.3) &40.7 (14.4)
    &$<$ 27.0     &64.0 (16.1)  &$<$ 30.1    \\
                     &83.9 Jy     &80.5 Jy        &72.7 Jy        &73.5 Jy
  &64.9                &63.8 Jy        &50.6 Jy        &38.0 Jy       \\
HD 36112  &$<$8.4      &7.9  (2.5)      &$<$ 8.2       &$<$12.4      &$<$ 11.4
     &$<$ 12.8     &14.1  (4.1)  &16.0 (4.7)  \\
                     &13.2 Jy     &12.8 Jy        &12.5 Jy        &12.3 Jy
  &12.2 Jy           &12.0 Jy        &11.0 Jy        &8.2 Jy         \\
HD 97048  &$<$ 15.3   &$<$21.5     &$<$19.5      &25.8 (7.1)    &32.4 (5.3)
 &$<$ 24.5     &49.1 (11.5) &$<$ 23.8   \\
                     &55.1 Jy     &53.3 Jy        &54.2 Jy        &58.2 Jy
   &58.8 Jy          &58.1 Jy        &54.3 Jy        &43.9 Jy       \\
HD 100546&65.0 (8.8) &49.9 (5.7)   &64.7 (8.5)    &71.5 (6.9)    &73.9 (10.4)
 &58.8 (9.7)    &88.3 (9.7)   &60.0 (8.3)   \\
                     &75.1 Jy     &68.4 Jy        &64.1 Jy        &60.5 Jy
  &57.0 Jy           &52.8 Jy        &45.9 Jy       &35.0 Jy        \\
IRS 48        &9.9 (3.0)    &12.3 (4.1)  &11.1 (3.7)   &13.4 (2.8)     &22.7
(4.7)      &12.0: (4.8)    &$<$ 18.4    &$<$ 21.5\\
                     &18.7 Jy     &17.0 Jy       &15.7 Jy       &14.4 Jy
 &13.3 Jy           &12.3 Jy         &10.2 Jy       &7.2 Jy\\
AS 205        &22.1 (2.3) &20.3 (2.8)   &21.6 (2.4)   &18.5 (2.6)    &18.9
(3.9)       &13.0  (3.8)   &24.0 (5.5)  &16.3 (4.2)   \\
                     &13.1 Jy      &12.8  Jy      &12.7 Jy       &12.7 Jy
  &12.8 Jy           &12.6 Jy        &11.9 Jy       &9.1 Jy \\
S CrA          &34.1 (2.5)  &26.7 (2.7)  &34.2 (2.9)   &44.2 (3.9)     &44.7
(4.3)      &52.4 (4.7)    &50.5 (7.5)  &39.7 (6.6)\\
                     &16.9 Jy        &16.1 Jy       &15.7 Jy       &15.3 Jy
     &15.0 Jy          &14.7 Jy        &13.4 Jy       &9.8 Jy\\
DG Tau       &46.4 (2.0) &42.5 (2.0)  &45.9 (2.3)   &50.0 (2.9)    &47.7 (2.9)
     &44.9 (3.5)     &67.7 (4.5)  &31.0 (3.9)\\
                     &17.6 Jy      &16.9 Jy      &16.8 Jy       &16.5 Jy
&16.2 Jy            &15.9 Jy         &14.4 Jy       &10.4 Jy\\
RU Lup       &$<$ 9.0     &10.8 (2.9)  &11.5 (2.1)   &--&--&--&--&--\\
                     &4.2 Jy        &4.0 Jy         &4.1 Jy
&--&--&--&--&--\\
                     \hline
\end{tabular}
\end{center}
\label{t_lineflux}
\tablefoot{\tablefoottext{a}{
The transition $J$ = 31 \ra \ 30 is a  blend with OH, so the values listed are upper limits. 
The transitions $J$ = 26 \ra \ 25 at 100.46 \mic \ and $J$ = 13 \ra \ 12 at 200.41 \mic,
are not listed, as the data are too noisy to give a meaningful upper limit.}}
\end{table*}

%++++++++++++++++++++++++++++++++++++++++++++++++++++++++++++++++++++++
\section{Results}
\label{s_res}

\subsection{Detections}

We detected CO lines in only 5 HAEBE stars of our sample: AB Aur, HD 36112, HD 97048, HD 100546, and 
IRS 48. While all 5 detections toward HAEBEs occur for fairly nearby stars $\mean{d} = 159\pm71$ pc, there 
are many HAEBEs at similar distances without detections. All 5 HAEBEs with detections have group I (flaring) 
discs; there is no CO detection seen in a group II (flat) HAEBE disc. Within the group I sources, the sources
with CO detections have the highest \luv \ observed.

For the TTS, we detected CO in 4 objects: AS 205, a disc wind source (Bast et al. \cite{bast2011}), and a possible 
outflow source (Mundt \cite{mundt1984}); S CrA and B: disc wind sources (Bast et al. \cite{bast2011}), RU Lup: a 
disc wind source (Bast et al. \cite{bast2011}), as well as an outflow source (Herczeg et al. \cite{herczeg2005} and 
references therein); and DG Tau: an outflow source (e.g. Lavalley-Fouquet et al. \cite{lavalley2000}). An equivalent 
disc geometry classification as for the HAEBEs does not exist, so we could not compare the two samples in this regard. 
However, CO detections are only seen in TTS that show evidence for a strong disc wind (according to the line 
shape criterion defined in Bast et al. \cite{bast2011}), or a resolved outflow, or both. It is not yet known whether there is 
a correlation between the presence of disc winds and outflows, although several discs have signatures of both. 

The highest observable transition is $J$ =  49 \ra \ 48, but the highest detected transition in our spectra 
is $J$ = 36 \ra \ 35, seen only in HD 100546 and the outflow source DG Tau. In Table~\ref{t_transitions}, 
we list the transitions, their wavelength and the upper level energy (in kelvin) that are detected in our spectra.
The transition $J$ = 31 \ra \ 30 is at the same wavelength as an OH line; therefore we consider it to be an upper 
limit in our further analysis.

\subsection{Rotation Diagrams}

\begin{figure*}
\begin{center}
   \resizebox{8cm}{5.2cm}{\includegraphics {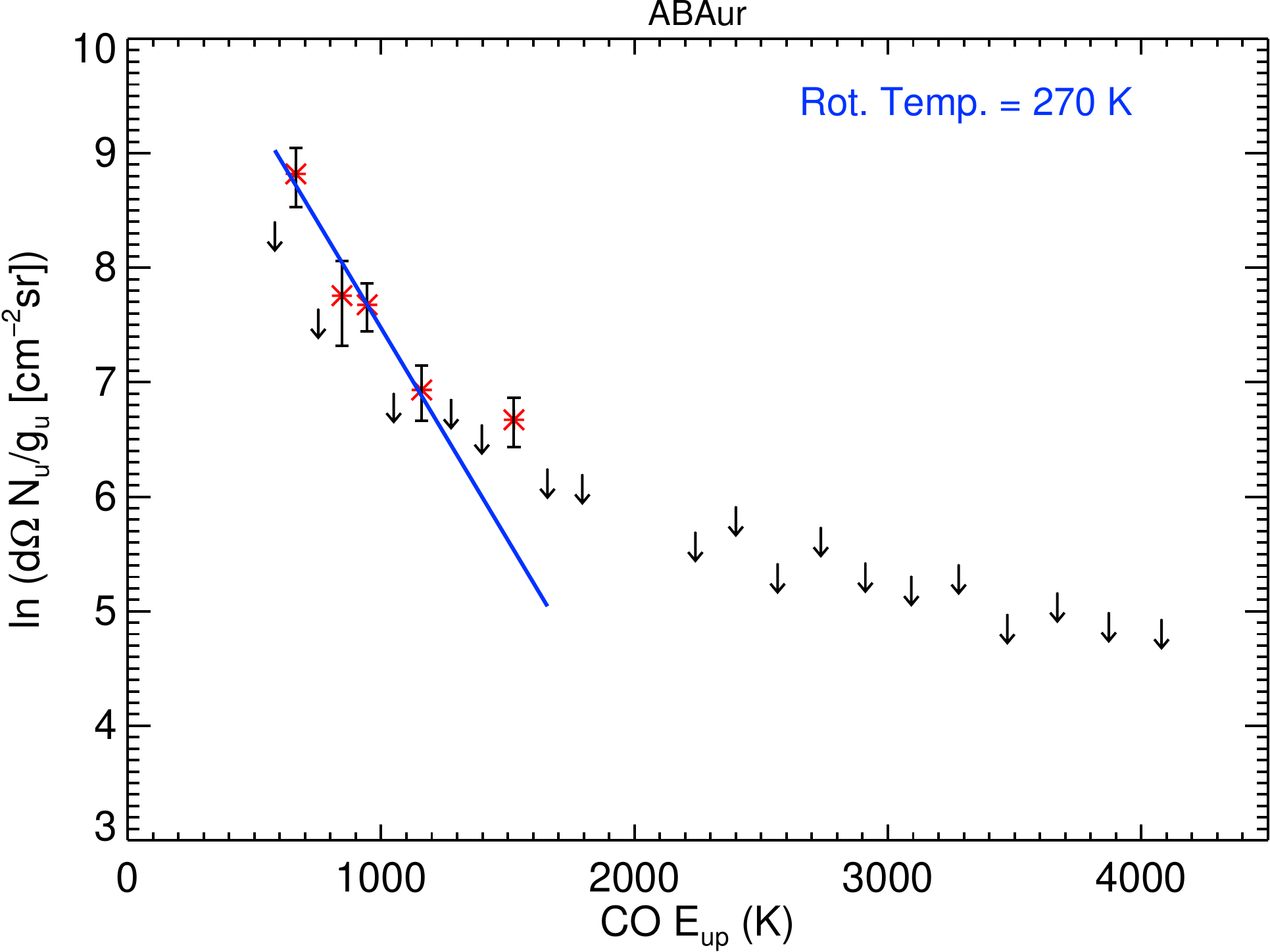}}
   \resizebox{8cm}{5.2cm}{\includegraphics {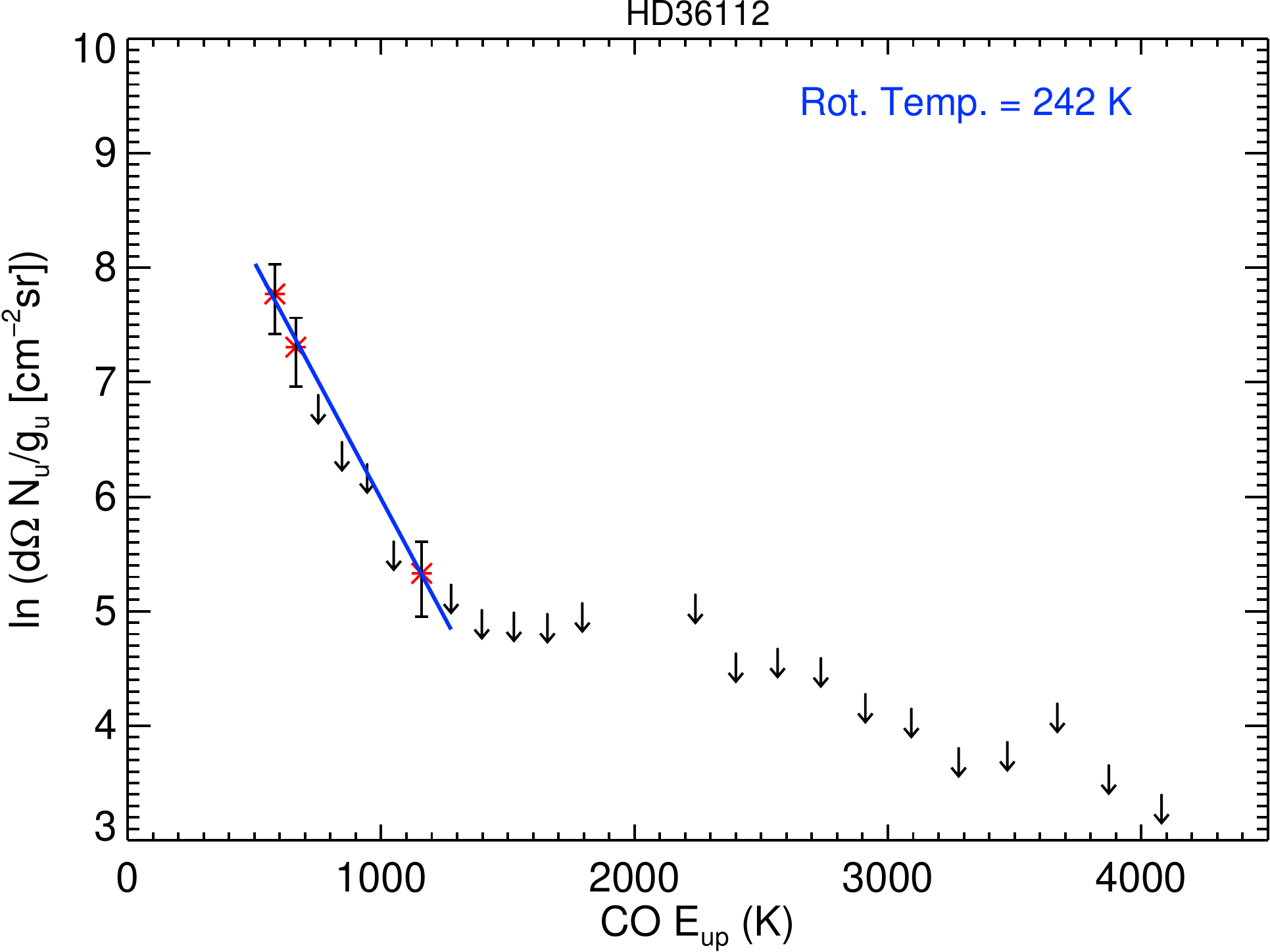}}
   \resizebox{8cm}{5.2cm}{\includegraphics {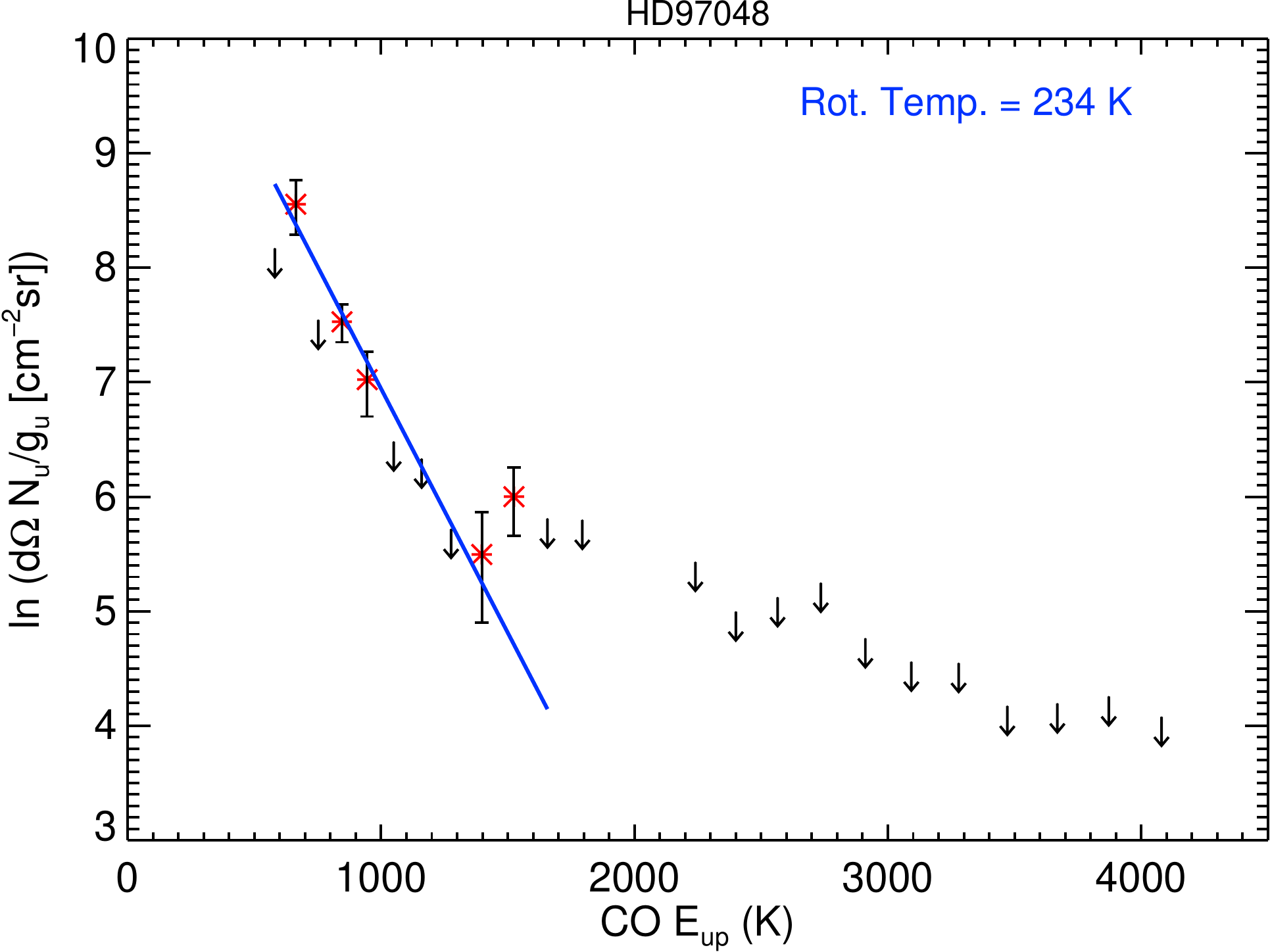}}
   \resizebox{8cm}{5.2cm}{\includegraphics {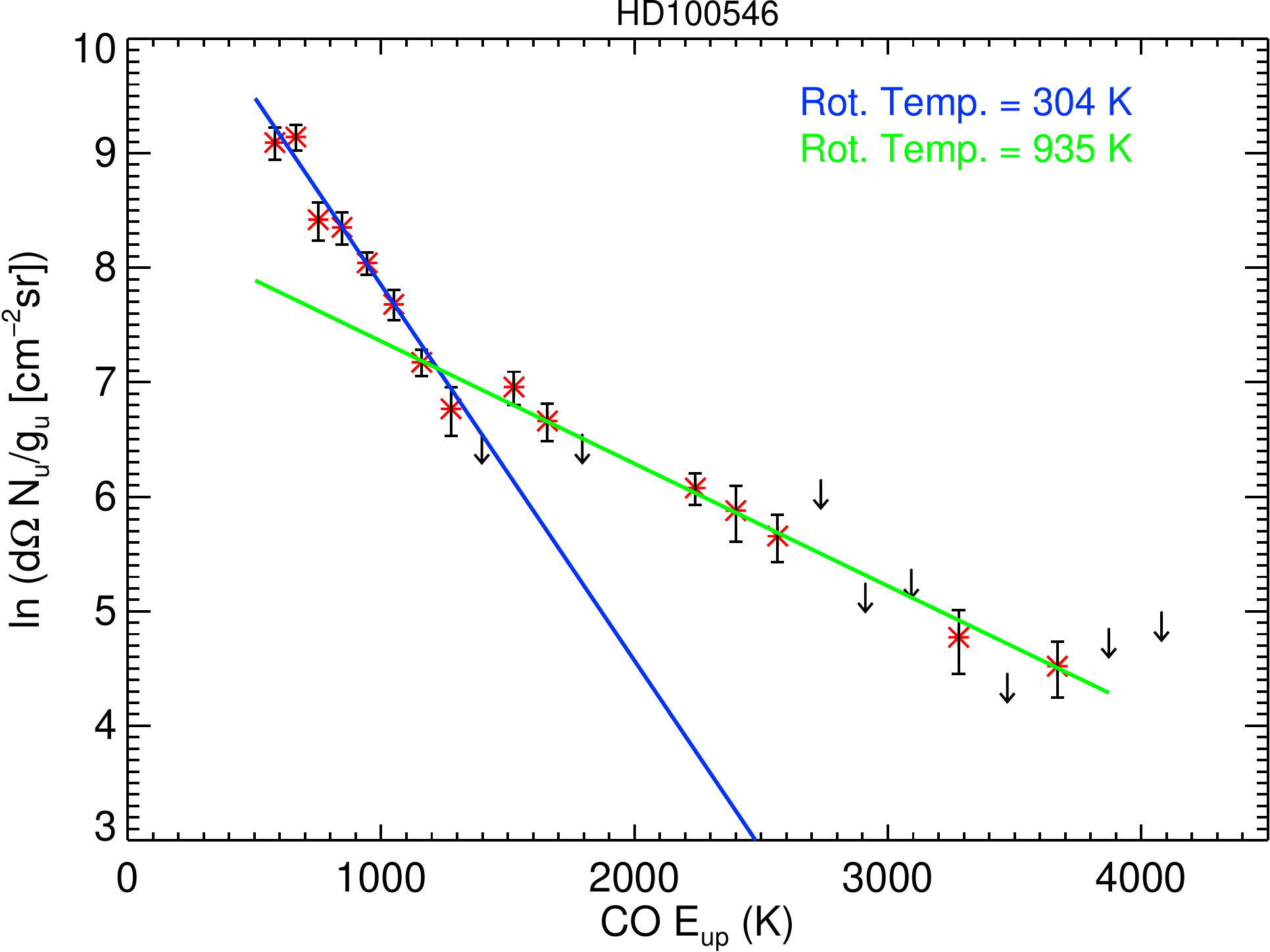}}
   \resizebox{8cm}{5.2cm}{\includegraphics {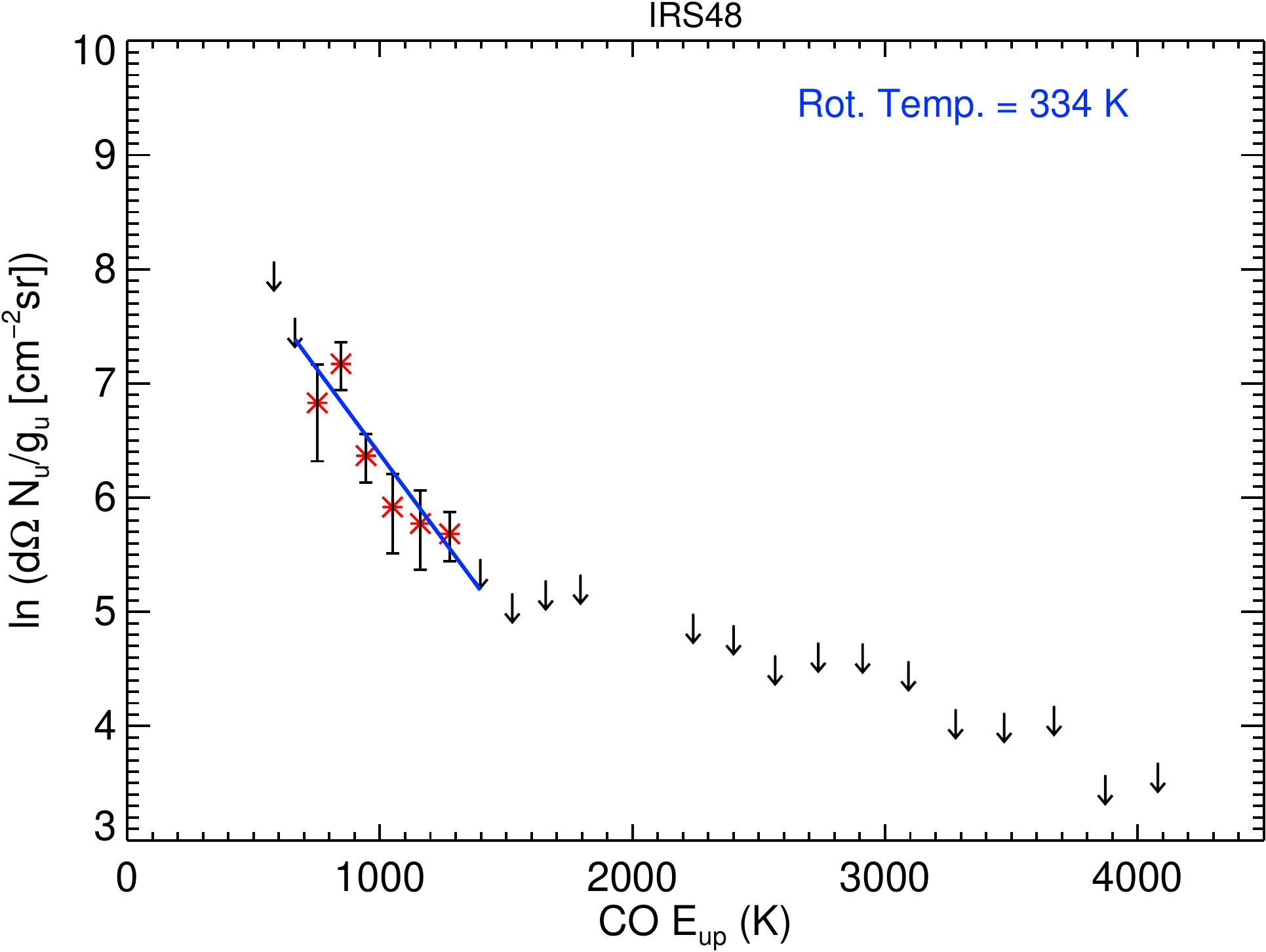}}
   \resizebox{8cm}{5.2cm}{\includegraphics {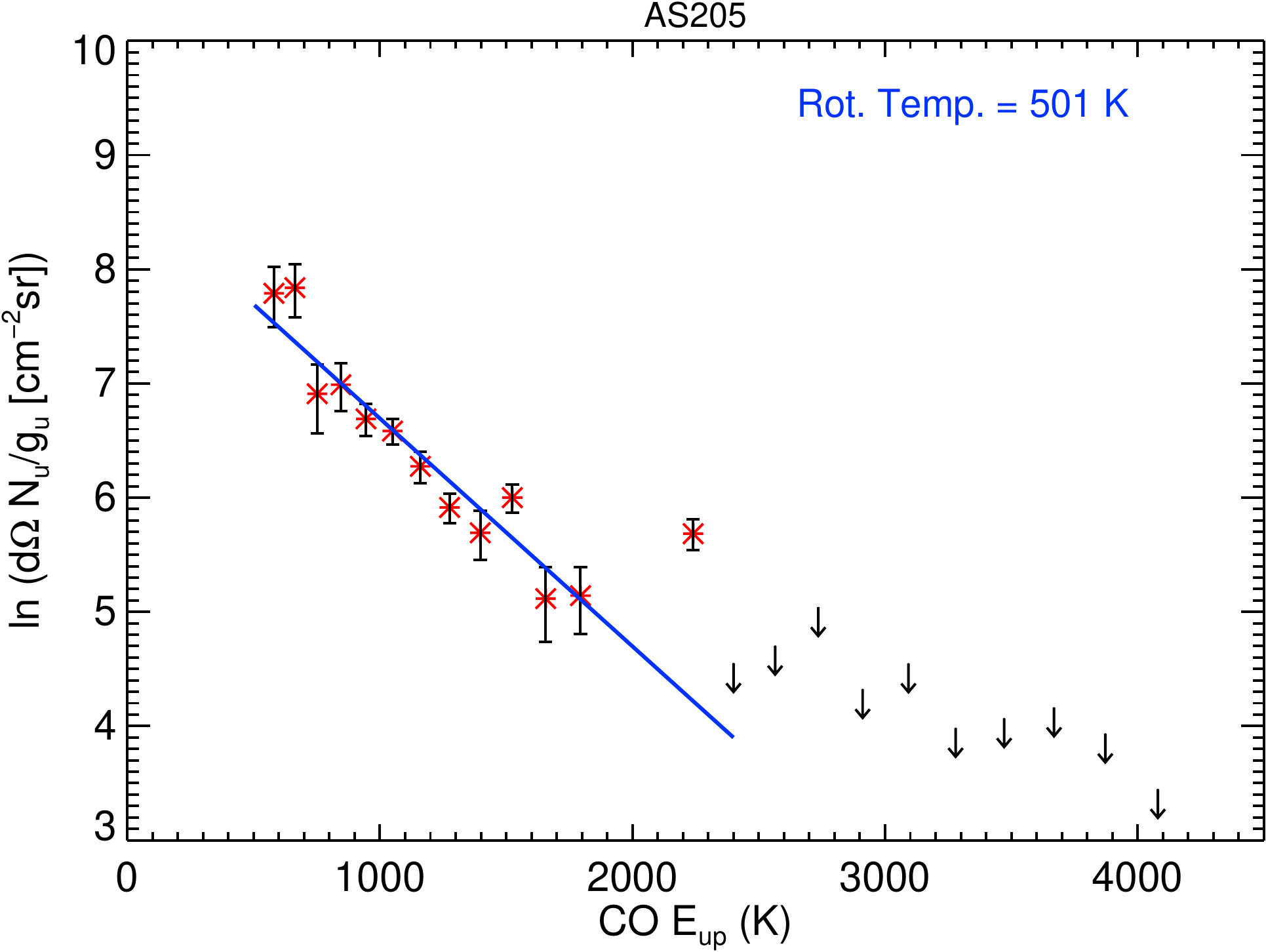}}
   \resizebox{8cm}{5.2cm}{\includegraphics {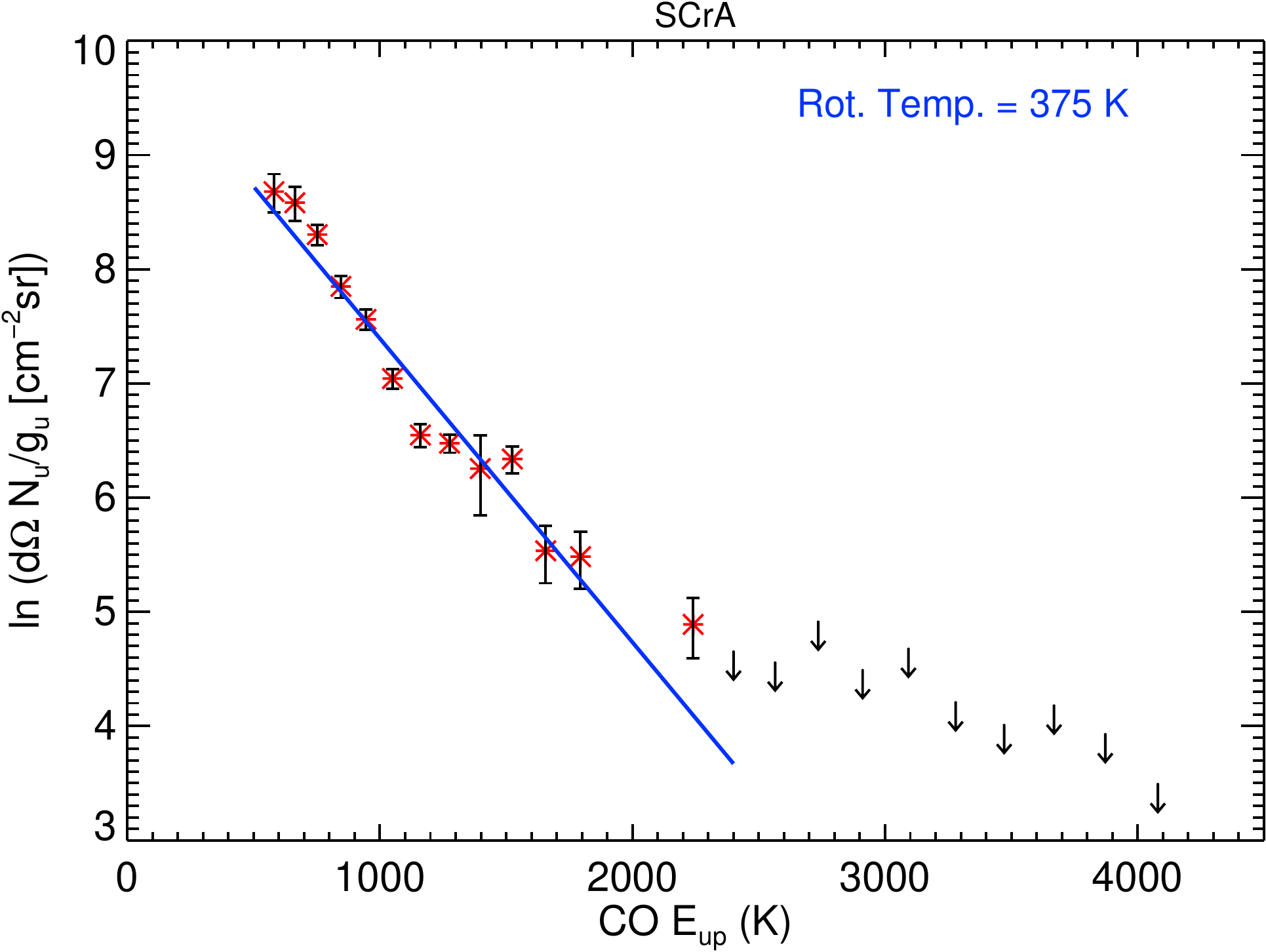}}
   \resizebox{8cm}{5.2cm}{\includegraphics {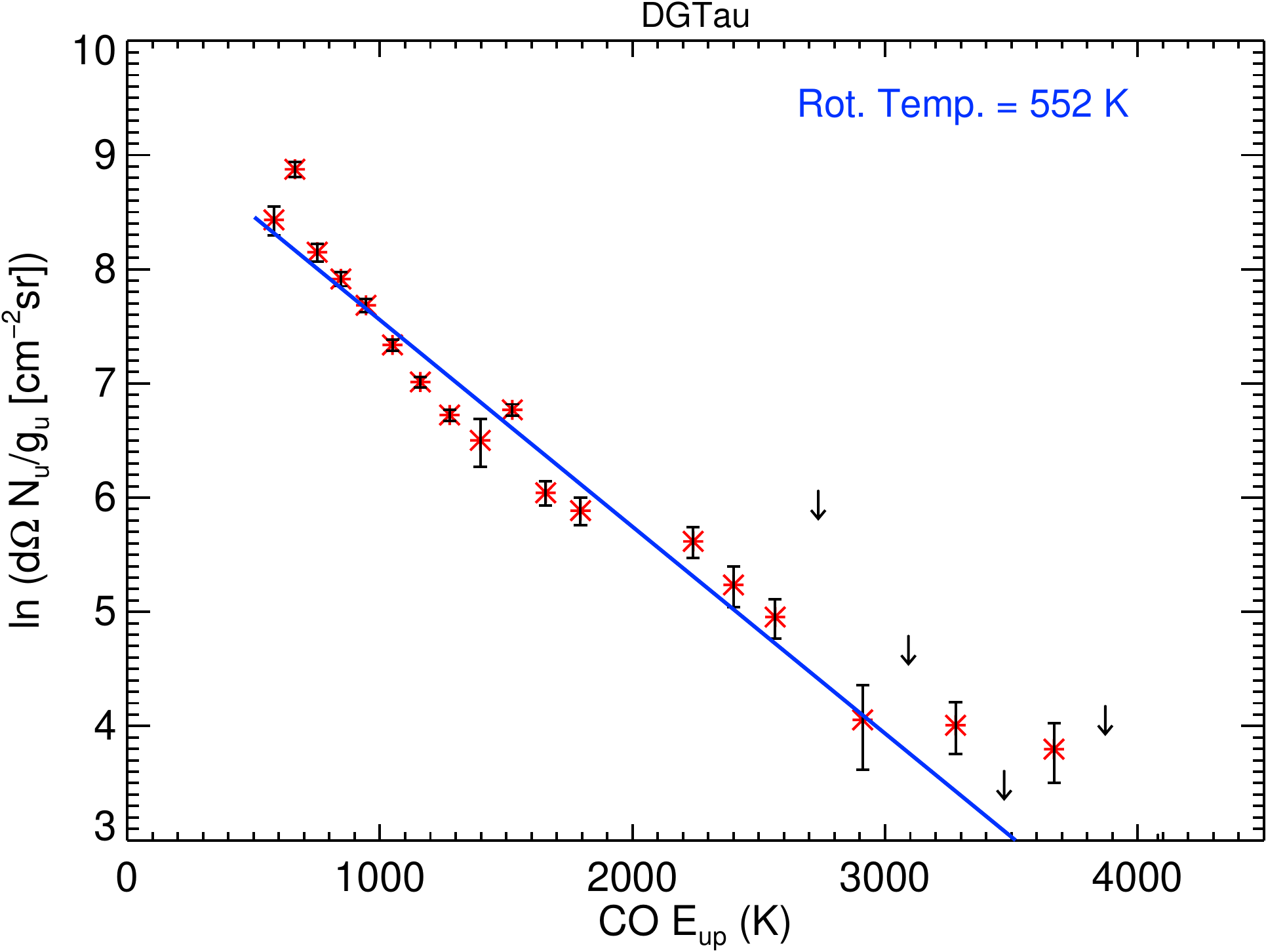}}
\caption{The rotational diagrams of the sources with at least 3 CO emission
line detections (marked with a red asterix, upper limits with an arrow). We also
show a fit to the ladder, from which the rotational temperature can be derived.  }
\label{f_rotdia}
\end{center}
\end{figure*}

Since we covered a wide range in transitions it is possible to analyse our data with the aid of rotational 
diagrams (see Goldsmith \& Langer \cite{goldsmith1999}). We constructed rotational diagrams for the sources 
with at least three CO line detections (see Fig.~\ref{f_rotdia}), and list the derived \trot \  in Table~\ref{t_trot}. 
We derived the error on \trot \ by calculating the 2 extreme temperatures that are still consistent with the 
errors on the line fluxes. We define the Y axis of the diagram as $\ln \ (d\Omega \ $N$_{\rm{u}}$ / g$_{\rm{u}})
 =  \ln \ (4 \pi \ $F$_{\rm{u}}$ / (A$_{\rm{u}}$ h$\nu$  g$_{\rm{u}}$))  such to include the solid angle $\Omega$ of 
the emission, since the CO ladder could emerge from different regions in the disc (see discussion in BR12). For 
the HAEBEs, we obtain \trot\  between 200 and 350 K. For HD 100546 we also need a second component to 
fit the rotational diagram, with \trot \ $\sim$ 900 -1000 K. Also for AB Aur and HD 97048 a second component 
could be present, but in these cases we do not have enough line detections to derive a temperature. More 
sensitive data could reveal higher $J$ emission lines also for these sources. The result for 
HD 100546 is similar to that found in an earlier analysis (Sturm et al. \cite{sturm2010}). 
We note that, for a few transitions (around 22 \ra \ 21, or \eup = 1397 K), the fit predicts a too high line flux 
when compared to the observations. It is not clear whether to attribute this to bad flux calibration of this wavelength 
range ($\sim$ 118 \mic) and/or due to the low quality of this spectral region. The mean of the HAEBE 
\trot\ values, excluding the hot component in HD100546, is $\langle\trot\rangle  = 271\pm 39$ K.  For TTS, the 
temperatures are higher - between 350 and 600 K, with the outflow source DG Tau having the highest \trot \ : 
582 $\pm$ 12\,K. The mean value is $486\pm 104$ K, significantly higher than the mean for the HAEBEs.

The rotational temperatures of 200-350 K in HAEBE discs are lower than those typically found from $^{12}$CO 
ro-vibrational fundamental ($v$ = 1 \ra \ 0) near-IR transitions (Salyk et al. \cite{salyk2011}). They are comparable, 
however, to the rotational temperatures derived from the $^{13}$CO ro-vibrational fundamental lines, which probe 
somewhat further out into the disc (Brown et al. \cite{brown2013}).
   
We note that converting \trot\ derived from a rotation diagram to a kinetic temperature is not straightforward, as it 
assumes optically thin emission. Even if the lines are thin, \trot \ equals \tgas\ only if the excitation is in thermal 
equilibrium with the gas, i.e., if the density is high enough. Indeed, Bruderer 
et al. (\cite{bruderer2012}) discussed that the CO ladder reflects emission from the disc atmosphere from a range 
of radii ($\sim$ 20--80\,AU) at different temperatures, and that the shape of the CO ladder can depend on several 
factors: 1) the gas temperature and density; 2) for the low-$J$ to mid-$J$ lines, optical depth effects; and 3) the 
amount of emitting molecules if the line is optically thin, or size of emitting region, if the line is optically thick. For 
HD 100546, Bruderer et al. (\cite{bruderer2012}) predicted that lines up to $J$ = 16 \ra \ 15 are optically thick. 
Assuming that the model for HD100546 is relevant for all our sources, the lines we observe are mainly optically thin 
so that the gas temperature and density are the main factors, assuming a normal CO abundance.

\begin{table}
\caption{\trot \ derived from the rotational diagrams.}
\begin{center}
\begin{tabular}{lll}
\hline\hline
Object & \trot (K)& Notes \\
\hline
AB Aur         & 270 $^{+53}_{-39}$ & Highest \eup \ not fitted, 2 components? \\
HD 36112   & 242 $^{+41}_{-31}$ & \\
HD 97048   & 234 $^{+36}_{-27}$  &  Highest \eup \ not fitted, 2 components? \\
HD 100546 & 276 $^{+19}_{-17}$ & Needs 2 temperature components \\
                      & 935 $^{+81}_{-69}$ & \\
IRS 48          & 334 $^{+71}_{-50}$ & \\
\hline
AS 205         & 501 $^{+41}_{-36}$ &  Highest \eup \ not fitted, 2 components? \\
S CrA           & 375 $^{+14}_{-12}$  & \\
DG Tau        & 582 $^{+12}_{-11}$ & \\
\hline
\end{tabular}
\end{center}
\label{t_trot}
\end{table}

%================================================================================
\section{Analysis}
\label{s_ana}

\subsection{Relation with the (UV) luminosity}

Does the far-IR CO line emission depend on stellar luminosities? We compared average HAEBE and 
TTS CO line luminosities for the $J$ = 17 \ra \ 16 transition, and found that they are remarkably similar:  
$(2.06\pm0.7) \times 10^{-5}$ \Lsun\ for the HAEBEs and $(2.06\pm1.0) \times 10^{-5}$ \Lsun\  for the TTS, 
while the average bolometric luminosities are 86.0 \Lsun \ and 5.0 \Lsun \ for the HAEBEs and TTS, 
respectively. This indicates that there is no correlation between the average line luminosity and \lbol.

But what about the UV luminosity? Models suggest that UV radiation is important for the gas heating and 
the excitation of the molecules, and one of the main factors determining the strength of the emission lines 
observed in the far-IR (e.g. Woitke et al. \cite{woitke2010}; Kamp et al. \cite{kamp2011}). Also BR12 showed 
the importance of the temperature of the stellar radiation field with respect to the amount of line flux (see their 
Fig. 13). Because of self-shielding, the CO molecule can survive the UV radiation in the surface layers and 
the CO emission lines trace the heated disc surface. Since CO is an important coolant, an increased amount 
of heating leads to the detection of higher $J$ lines, as well as higher line luminosity. 

HAEBE stars with strong \luv \ and flaring discs (group I) often have high \Odrie \ \mic \ line flux (Meeus et al. 
\cite{meeus2012}; paper I). Furthermore, sources with higher \Odrie \ fluxes also have higher far-IR CO fluxes, 
as is shown in Meeus et al. (\cite{meeus2012}) for the CO $J$ = 18 \ra \ 17 transition.

In HAEBEs, the bulk of the UV radiation is stellar, and the relative contribution from accretion is small. 
In contrast, the accretion luminosity is the most important contributor to the UV luminosity of TTS (e.g. 
Yang et al. \cite{yang2012}), as the relatively cool stellar photospheres produce very little UV flux.  
In Sect.\ref{s_res}, we showed that \trot \ is higher in our sample of TTS than in the HAEBEs. This can perhaps
be attributed to a difference in gas heating mechanism, which is stellar in HAEBEs, while it is due to accretion 
and could also be due to shocks in the TTS. Podio et al. (\cite{podio2012}) and Karska et al. (\cite{karska2013}) 
show that, in outflow sources, heating by shocks is necessary to explain the observed molecular line emission. 
In our TTS sample, at least two sources with CO detections (DG Tau and RU Lup) show evidence for a jet.

Given the differences in origin of the UV radiation, important in the context of gas heating 
mechanisms, we will discuss HAEBEs and TTS separately in the following sections.

\subsubsection{HAEBEs}

The five HAEBEs with CO detections all have flaring discs, with a hot disc atmosphere. Four out of five 
detections are in stars with a relatively high \luv, but there are also non-detections at \luv \ values similar 
to those of the detections, with the transitional disc HD 141569 A having the highest value. Unfortunately, the 
modest ranges in \luv\ and line luminosity, coupled with the small sample size, make it impossible to 
statistically test whether the line luminosity correlates with \luv.

\subsubsection{TTS}

\begin{figure}
   \resizebox{\hsize}{!}{\includegraphics {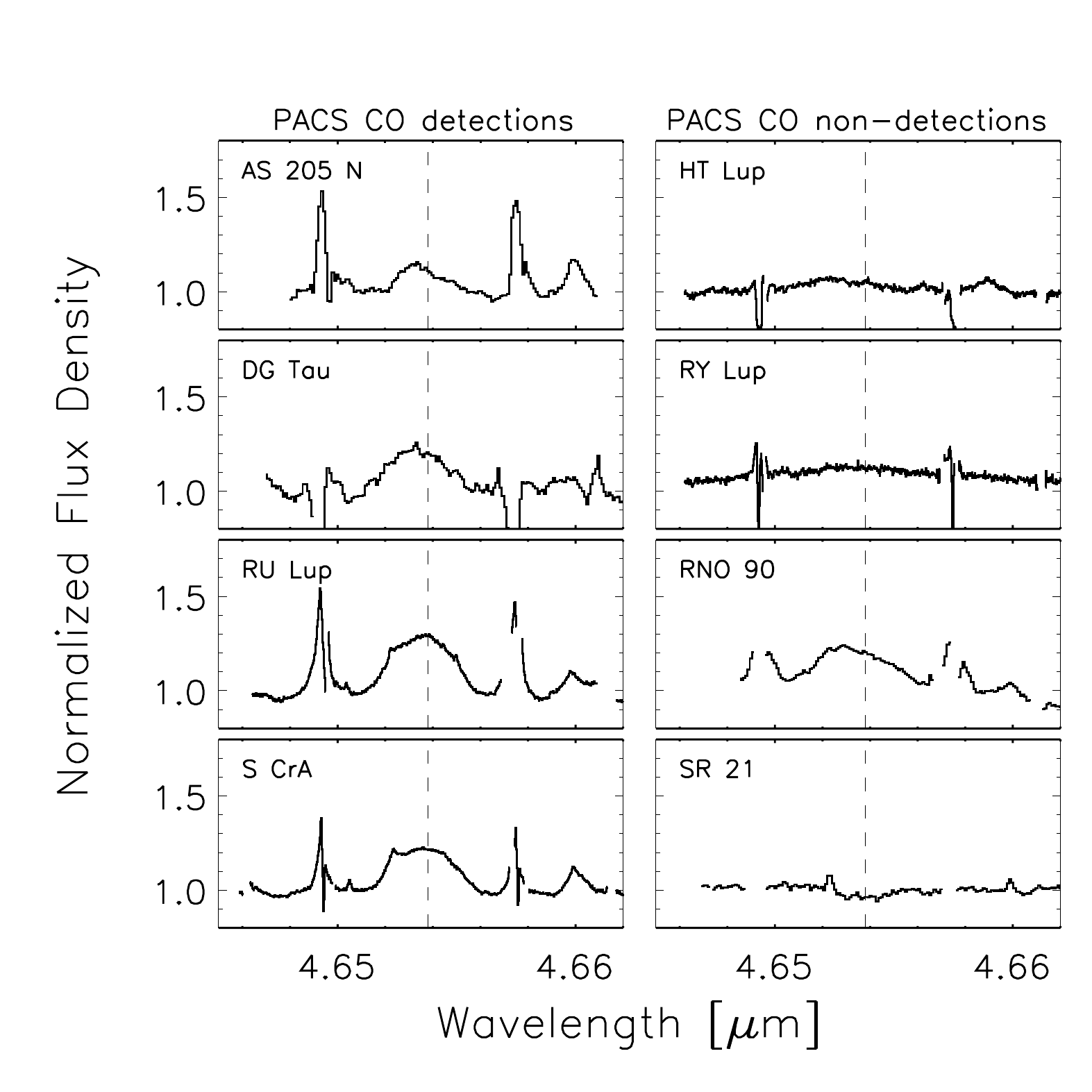}}
\caption{Observations of H Pf$\beta$ for the TTS sample, line position indicated with the vertical dashed line. 
Left: sources for which we have CO detections, right: sources for which CO was not detected with PACS. }
\label{f_pfbeta}
\end{figure}

Salyk et al. (\cite{salyk2013}) 
show that the accretion rate can also be correlated with the strength of the H Pf$\beta$ (n = 7 \ra \ 4) line at 4.65 \mic, 
which was observed in all of the TTS in our sample. In the following, we use data from that paper, taken with either 
VLT/CRIRES (Pontoppidan et al. \cite{pontoppidan2011}) or Keck/NIRSPEC. In Fig.~\ref{f_pfbeta} we plot the 
H Pf$\beta$ line emission both for sources that do show the PACS CO 
lines and those that do not. All the sources with CO detections have strong H Pf$\beta$ lines, while most of the sources 
without -- with the exception of RNO 90 -- have weak H Pf$\beta$ lines.  Also the ro-vibrational CO lines appear to be 
stronger in the sources with PACS detections, especially the $v$ = 2 \ra \ 1 R(8) line at 4.6598 \mic, which is enhanced 
by UV radiative excitation (Brown et al. \cite{brown2013}). This implies a connection between accretion rate and CO 
detections, perhaps due to the UV radiation provided by the accretion column.  However, a direct connection with the 
value of \lacc \  is not found. We see a similar spread in \lacc \  for the detections as the non-detections, so that there is 
no clear correlation between these variables.

\subsection{Thermo-chemical models of the mid- to high-$J$ CO emission.} 

To explain the trend that far-IR CO lines are only detected towards flared discs, we ran a grid of 
thermo-chemical models introduced by Bruderer et al. (\cite{bruderer2012}). These models solve for the 
dust radiative transfer, the chemistry and the thermal balance in a self-consistent way.
Details of the model and benchmark tests are reported in BR12. In that paper, the mid- to 
high-$J$ CO emission of HD 100546 is reproduced together with the fine structure lines of [\ion{O}{I}] 
for a model tailored specifically to that disc. Here, we run a grid of generic disc models using a 
parameterized density structure. The surface density profile $\Sigma (r)$ of the disc varies as 

\begin{equation}
  \Sigma(r) \propto r^q 
\end{equation}

while the vertical density structure $\rho(r,z)$ is a Gaussian 

\begin{equation}
  \rho(r,z) \propto \Sigma(r) \cdot \exp(-1/2(z/{\rm H}(r))^2)
\end{equation}

The scale-height H(r) is given by the power-law 

\begin{equation}
{\rm H}(r) = {\rm H}_0 \left( \frac{r}{100 {\rm AU}} \right)^{\psi+1}
\end{equation}

The parameters assumed in the grid are given in Table \ref{tab:thermparam}. We vary the disc mass 
$M_{\rm disc}$ and the flaring parameter $\psi=0.0, 0.1, 0.2$ to study the dependence of the 
CO lines on these two parameters.

As a word of caution, we note that thermo-chemical modeling is prone to large uncertainties, in particular 
in the gas temperature calculation (R\"ollig et al. \cite{rollig2007}). We thus refrain from fitting individual 
objects, but aim to reproduce the main trends found in the observations.

\begin{table}[tbh]
\caption{Parameters employed in the thermo-chemical models}
\label{tab:thermparam}
\centering
\begin{tabular}{lll}
\hline\hline
Parameter &  Value &  \\
\hline
Disc mass                 & $M_{\rm disc}= 0.1, 10^{-2}, 10^{-3}, 10^{-4}$ $M_\odot$\\
Flaring angle             & $\psi=0.0, 0.1, 0.2$ \\
Inner radius              & $r_{\rm in}=0.383$ AU\tablefootmark{a} \\
Outer radius              & $r_{\rm out}=400$ AU\\
Scale height at 100 AU    & H$_{100\, {\rm AU}}=10$ AU\\
Surface density power-law & $q=-1$ \\
Stellar luminosity        & $L_*=30$ L$_\odot$ \\
Stellar temperature       & $T_*=10^4$ K \\
Stellar X-ray luminosity  & $L_{\rm X}=10^{29}$ erg s$^{-1}$ \\
Gas-to-dust mass ratio    & g/d$=$100 \\
Distance                  & $d=$150 pc \\
Inclination               & $i=30^\circ$ \\ 
Grain size                & $a=0.1-10$ $\mu$m \\
Grain size distribution   & $dn/da \propto a^{-3.5}$\\   
Grain composition         & ISM (Weingartner \& Draine \cite{weingartner2001}) \\
\hline
\end{tabular}
\tablefoot{\tablefoottext{a}{Dust sublimation radius, $0.07 \sqrt{L_*/L_\odot}$ AU 
(Dullemond et al. \cite{dullemond2001}).}}
\end{table}

\begin{figure*}[!htb]
\center
\includegraphics[width=1.0\hsize] {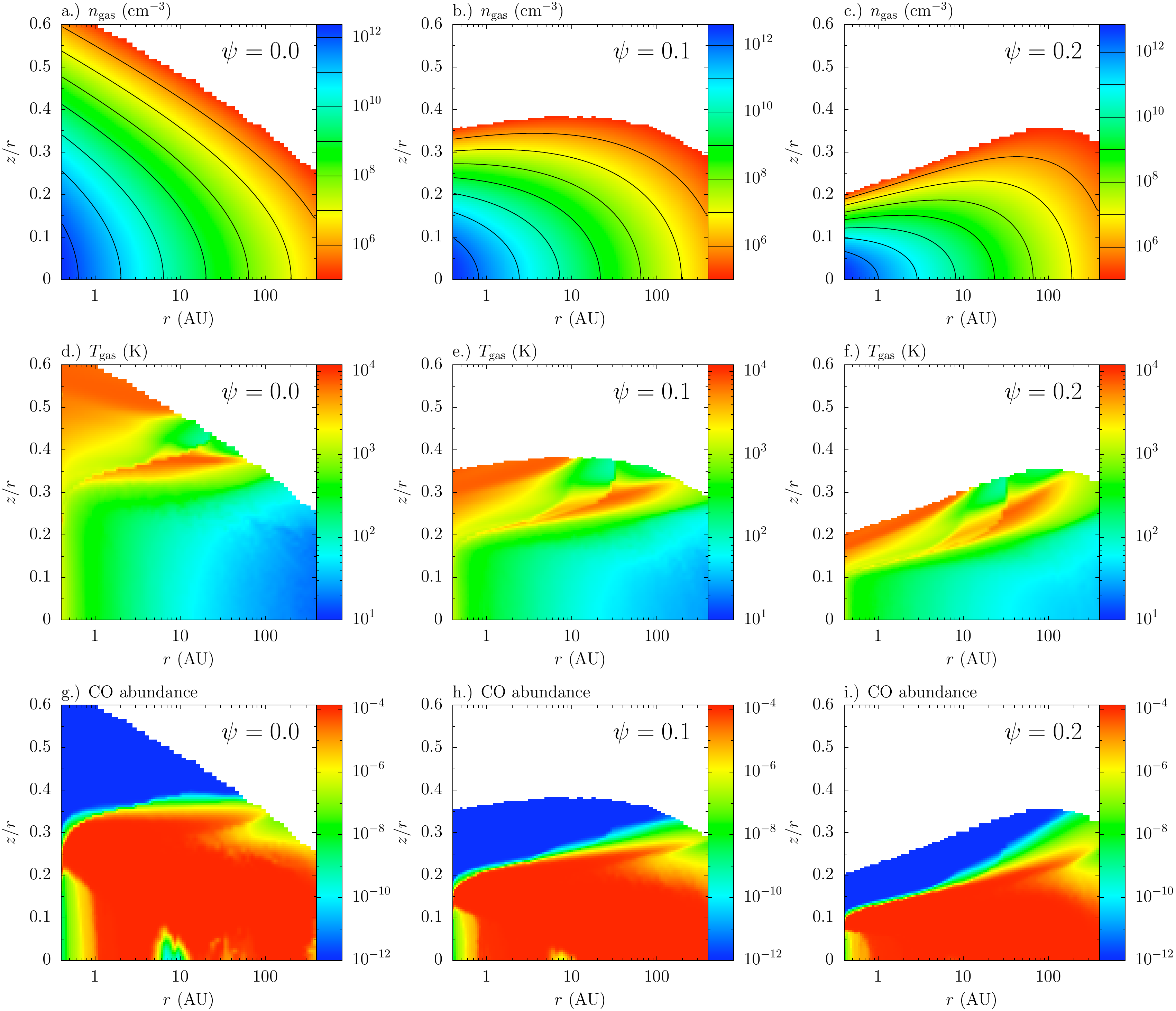}
\caption{Model density ($n_{\rm gas}$, top panels) gas-temperature ($T_{\rm gas}$, center panels) and 
CO abundance ($n_{\rm CO}/n_{\rm gas}$, bottom panels) of thermo-chemical models with 
$M_{\rm disc}=10^{-2}$ $M_\odot$ and different flaring parameter $\psi$=0.0, 0.1 or 0.2 (left to right columns).}
\label{fig:struct}
\end{figure*}

How does the flaring parameter affect the temperature structure? In Figure \ref{fig:struct}, the density, 
gas-temperature, and CO abundance profile of models with varying flaring parameters, $\psi=0.0, 0.1$ 
and $0.2$, and $M_{\rm disc} = 10^{-2}$ $M_\odot$ are shown. The figures are plotted such that the axes
are distance-normalised height ($z/r$) versus $\log(r)$. 
In the model without flaring ($\psi=0.0$) there is always an attenuating column of gas and dust 
between the star and regions with density $\gtrsim 10^6$ cm$^{-3}$. This density corresponds to the 
critical density of the CO $J=16-15$ transition ($E_{\rm up} = 750$ K), needed to excite the line collisionally. 
The mass of gas which has a temperature high enough (T $>$ 750 K) to excite the PACS CO lines, 
$M_{\rm {CO ex}}$, is -- in any of the models -- small compared to the total disc mass ($< 0.5 \%$). However, 
for larger flaring parameters, more gas can be irradiated and heated directly by the star to temperatures high 
enough for the excitation of the high-$J$ CO lines: 
for $\psi=0.0$, $M_{\rm {CO ex}}$ = $4.2 \times 10^{-6}$ $M_\odot$, for $\psi=0.1$, $M_{\rm {CO ex}}$ = 
$2.0 \times 10^{-5}$ $M_\odot$ and for $\psi=0.2$, $M_{\rm {CO ex}}$ = $4.5 \times 10^{-5}$ $M_\odot$; 
an increase up to a factor 10 with increasing flaring. Furthermore, the radial extent of the warm region also 
increases with amount of flaring. This is of importance, since BR12 find that lines up to $J=16-15$ can be 
optically thick so that the size of the emitting area determines the amount of line flux.

The abundance of CO (Figure \ref{fig:struct}g, \ref{fig:struct}h and \ref{fig:struct}i) is dominated by 
photodissociation in the upper atmosphere and the inner rim of the disc. The ``warm-finger'' of CO is 
produced by CO formation through the reaction of C$^+$ with H$_2$ to CH$^+$ which only proceeds at 
high gas temperature (Jonkheid et al. \cite{jonkheid2007}; Ag\'undez et al. \cite{agundez2008}). Starting from 
CH$^+$, a chain of reactions through CO$^+$ and HCO$^+$ then leads to CO. CH$^+$ is detected in both 
HD 97048 and HD 100546 (Thi et al. \cite{thi2011}; paper I). The CO lines discussed in this section form at 
large heights in the disc atmosphere and are not affected by any chemistry lower in the disc. 

\begin{figure*}[!htb]
\center
\includegraphics[width=0.9\hsize] {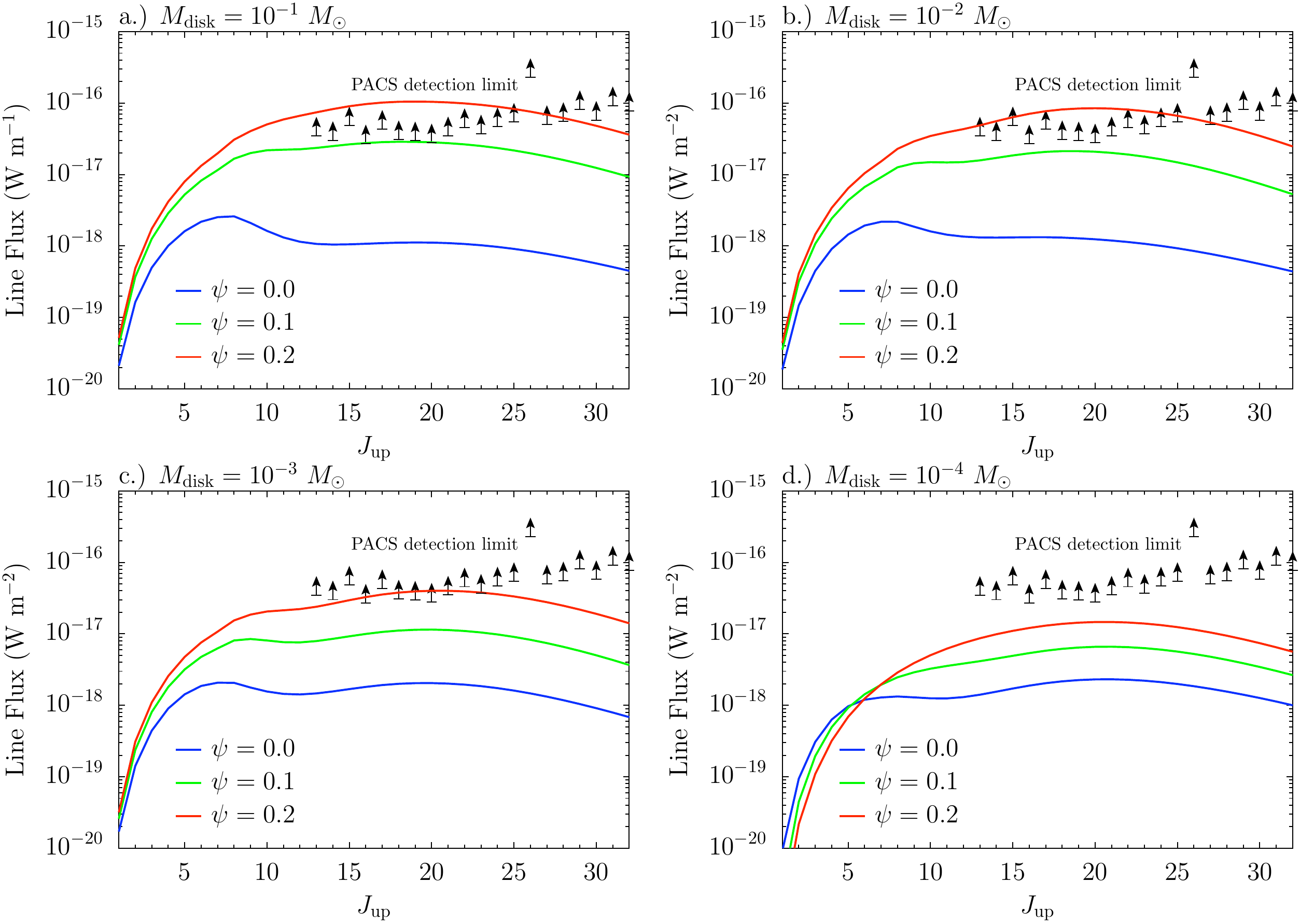}
\caption{Model CO ladder for the models with different gas mass $M_{\rm disc}$ and flaring parameter $\psi$. 
The lines with $J_{\rm up} \leq 7$ are convolved to the APEX beam at the appropriate frequency, the higher 
$J_{\rm up}$ to PACS. Black arrows indicate the typical PACS detection limit, extracted from the AB Aur spectrum.}
\label{fig:co_model_ladder} 
\end{figure*}

The CO ladder derived from the grid of models is shown in Figure \ref{fig:co_model_ladder}. Lines with 
$J_{\rm up} \leq 7$ are convolved to the beam of the APEX telescope which has been used to detect these 
lines towards Herbig discs (e.g. Pani\'c et al. \cite{panic2010}). Higher $J_{\rm up}$ lines are convolved to 
the beam of the PACS instrument. Typical PACS detection limits are shown by the example of AB Aur. We 
find that indeed the flux in the PACS lines depends considerably on the flaring of the disc. This is a direct 
result of the increasing mass of several 100 K warm gas with the flaring parameter $\psi$. 

On the other hand, the shape of the CO ladder does not vary significantly with disc mass, as long as $M_{\rm disc} > 
10^{-3}$ $M_\odot$. While the CO lines above $J_{\rm up} \sim 16$ are optically thin, the equilibrium of gas 
heating and cooling processes yields similar amounts of gas warm enough to excite the CO lines. This result 
is different from the findings of BR12, because they leave the dust structure unchanged and vary the 
gas-to-dust ratio. In the models presented here, we vary both gas and dust mass and keep the gas-to-dust 
ratio constant. It is the amount of dust which absorbs the impinging stellar light that is key for the heating of the gas. 

We conclude from this grid of thermo-chemical models, that the structure of the disc, and in particular the flaring 
which allows direct stellar irradiation of a larger volume of gas, is key for the understanding of the CO ladder observed 
by PACS. Non-flared discs produce too low line emission flux to be detected in our spectra.

%++++++++++++++++++++++++++++++++++++++++++++++++++++++++++++++++++++++
\section{Discussion} 
\label{s_disc}

\begin{figure}
   \resizebox{\hsize}{!}{\includegraphics {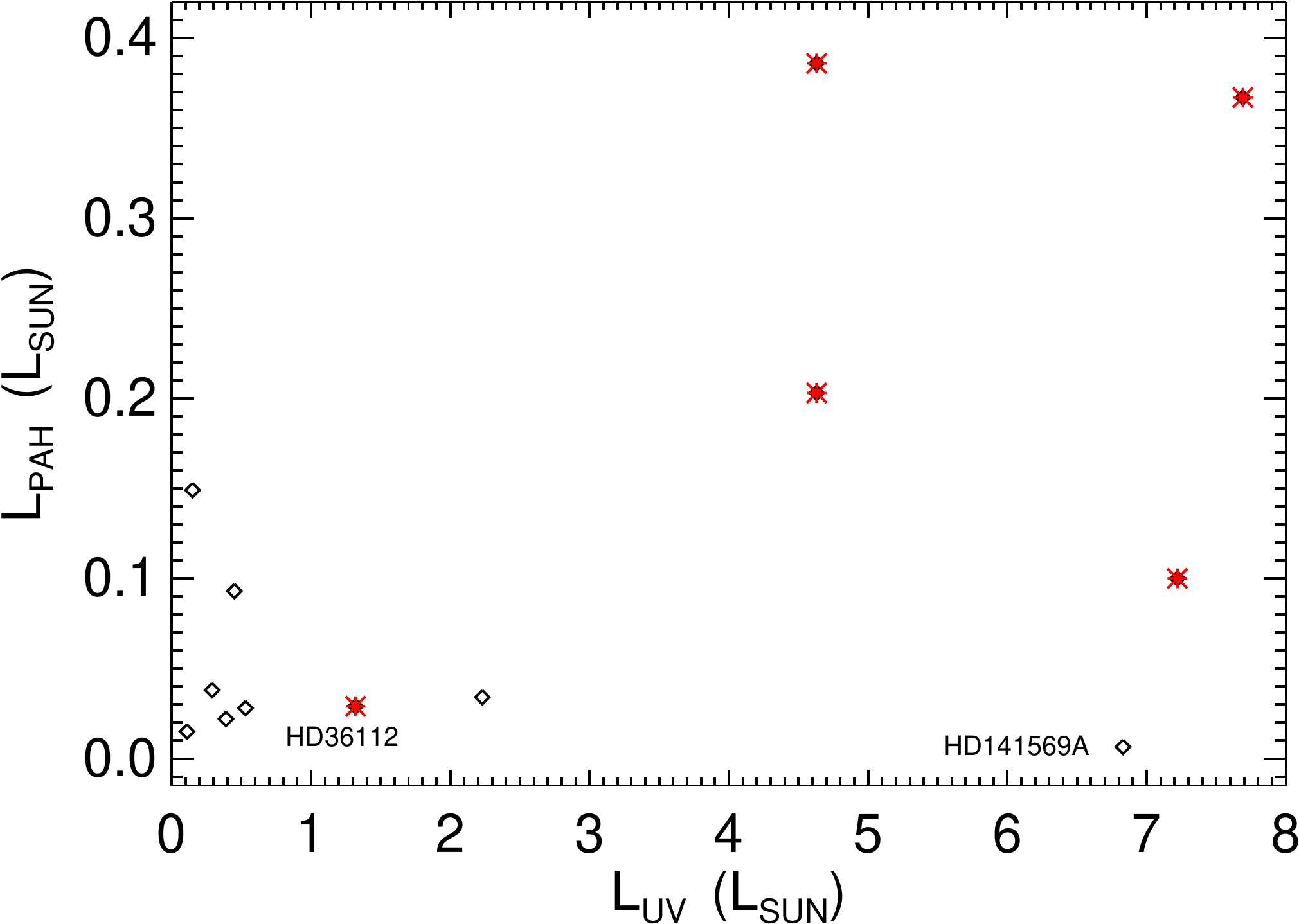}}
\caption{ \lpah \ as a function of \luv \ for the HAEBE sample. The red asterisks indicate
sources where CO was detected in our spectra, the black diamonds the sources with non-detections.}
\label{f_pahluv}
\end{figure}

What other parameters can influence the strength of the CO emission? In what follows, we speculate
about the influence that the PAH luminosity and the inner disc clearing can have on the far-IR CO emission.
While PAHs are frequently detected in HAEBEs (up to 70\%; Acke et al. \cite{acke2010}), they are seen in at 
most 15\% of T Tauri discs (Geers et al 2006; Oliveira et al. 2007, 2010); therefore we concentrate our 
discussion on PAHs observed in HAEBEs.

\subsection{PAH luminosity}

When present, PAHs contribute significantly to disc heating, absorbing UV photons in the upper disc layer 
after which they heat the disc through the photoelectric effect; as such they also have an influence on the 
observed line strengths (e.g. Kamp \& Dullemond \cite{kamp2004}; Bruderer et al. \cite{bruderer2012}).
Therefore, we searched for a link between the PAH luminosity and the observed CO emission. 
If we consider the sources with the highest \luv \ ( $>$ 4 \Lsun; AB Aur, HD 97048, HD 100546 - group I, 
and HD 141569 A, HD 150193 and HD 163296 - group II), we see that it contains equal amounts of sources
with flaring and non-flaring discs, but the group II sources lack CO far-IR emission. As the amount of 
\luv \ is not solely determining the strength of the CO lines, we now also take the PAH emission into account.
In Table~\ref{t_para}, we list \lpah \ for those objects where PAH bands are detected, while in Fig.\ref{f_pahluv}
we plot the \lpah \ as a function of \luv \ for both sources with and without CO detections. The group II sources 
HD 150193 and HD 163296 do not show PAH emission features, while the transitional disc HD 141569 A only 
has weak PAH bands when compared to the group I sources: \lpah $~\sim $ 0.007 \Lsun versus 0.030 to 0.39 \Lsun \
for the 5 HAEBEs with CO detections. This suggest that strong PAH emission, in addition to strong 
UV luminosity, enhances CO far-IR emission, consistent with the non-detection in HD 141569 A. However,  
the PAH luminosity does not uniquely determine the CO line flux. It is interesting in this
context to mention studies of the ro-vibrational bands of CO at 4.7~\mic, revealing that HD 97048, HD 100546
and HD 141569 A have UV fluorescence, so that \tvib $> $ 5600 K, while their \trot \ is distinct: $\sim$ 1000 K
for the first two, and $\sim$ 250 K for HD 141569 A (van der Plas, Ph.D.; Brittain et al. \cite{brittain2007}), 
again indicating that the PAH luminosity is an important factor in the context of gas heating. However,
BR12 showed that, in their model of HD 100546, a change in the PAH abundance changes the high-$J$ flux
by less than a dex, leaving flaring as the most important parameter determining CO line fluxes.

\subsection{Inner disc clearing}

Based on a small sample of group I sources with spatially resolved mid-IR images, Maaskant et al. 
(\cite{maaskant2013}) suggest that group I sources have dust-depleted regions in the inner disc 
($\lesssim$50 AU) that let the radiation penetrate further out in the disc. As a result, the outer regions, most notable 
the inner edge of the outer disc, experience more heating causing an excess at mid- to far-IR wavelengths.
On the other hand, no flat (group II) discs have reported gaps in the literature. This suggests that these discs are 
relatively more self-shadowed and consequently cooler in the outer parts of the disc. In our HAEBE sample with 
CO detections, HD 97048, HD 100546, IRS 48, HD 36112 and AB Aur all have dust-depleted inner regions 
(Maaskant et al. \cite{maaskant2013} for HD 97048; Bouwman et. al. \cite{bouwman2003} for HD 100546; Geers et al. 
\cite{geers2007}, Brown et al. \cite{brown2012} and van der Marel \cite{marel2013} for IRS 48; Isella et al. 
\cite{isella2010} for HD 36112; Honda et al. \cite{honda2010} for AB Aur). This indeed suggests that inner
dust clearing can result in more efficient heating of the outer layers, so that the amount of warm CO is increased.

\subsection{Summary}
The discussion above suggests that the CO line luminosity in HAEBEs can, besides dust and gas temperature 
decoupling in the surface layers (due to flaring) and high UV luminosity, also be enhanced by a combination of 
1) PAH heating and 2) enhanced UV penetration of the outer disc due to a dust-depleted inner region. We expect 
these factors to also be inter-linked, but a detailed parameter study of disc models would be needed to explain the causality.

%++++++++++++++++++++++++++++++++++++++++++++++++++++++++++++++++++++++
\section{Conclusions}
\label{s_conc}

We analysed PACS spectra covering 53.5 to 190 \mic \ to study the far-IR CO ladder
in a sample of HAEBE and TTS. Our results can be summarised as follows:

\begin{enumerate}

\item{Out of 22 HAEBEs and 8 TTS, we detected CO emission lines in only 5 HAEBE (23\%) and 4 TTS (50\%). }

\item{In HAEBEs, the far-IR CO lines are only detected in flaring discs, indicating that the disc geometry
(presence of a hot disc atmosphere) is important in creating detectable line emission. TTS with detections 
have evidence for a disc wind and/or outflow.}

\item{The highest $J$ transition ($J$ = 36 \ra \ 35), with \eup \ =  3669 K) was observed in HD 100546 and the 
outflow source DG Tau.}

\item{We constructed rotational diagrams for the objects with at least 3 CO detections, and derived \trot \  
between 200 and 500 K. The mean values of the warm components are $\langle \trot \rangle = 271\pm 39$ 
K for the HAEBEs and $\langle \trot \rangle = 486\pm 104$ K for the TTS. The HAEBE HD 100546 is the only 
object that clearly needs a second, hot component of $\sim$ 900 K, but other objects might also reveal a
hot component when observed with a more sensitive instrument. }

\item{We analysed the observed line fluxes with a small grid of thermo-chemical models. We found that an 
increased amount of flaring increased the CO line flux; a certain amount of flaring is necessary for the lines
to become detectable with PACS and can increase the line flux by a factor 10. This is a direct result of the
increase in warm gas mass with increasing amount of flaring where  \tgas \ and \tdust \ are decoupled. 
Furthermore, we found that the disc mass is not a sensitive parameter, as long as it is higher than a few 
10$^{-4}$ \Msun.}

\item{For the HAEBEs, amongst the group I sources, the ones with highest \luv \ have detectable CO lines.}

\item{In TTS, sources with strong Pf-$\beta$ lines (a tracer of accretion rate) tend to have CO emission, perhaps 
because of the UV heating provided by the accretion column.  However, a quantitative connection between \lacc \ 
and CO detections is not yet clear.}

\item{We suggest that both strong PAH luminosity and the presence of a dust-depleted inner disc region can 
increase disc heating, and by consequence, enhance the far-IR CO line luminosities.  }

\end{enumerate}

%++++++++++++++++++++++++++++++++++++++++++++++++++++++++++++++++++++++
\begin{acknowledgement}

We would like to thank the PACS instrument team for their dedicated support. G. Meeus is supported by
Ramon y Cajal grant RYC-2011-07920. B. Montesinos is partly supported by AYA-2011-26202.
Support for this work, part of the Herschel 'Open Time Key Project' Program, was provided by NASA through
an award issued by the Jet Propulsion Laboratory, California Institute of Technology.
PACS has been developed by a consortium of institutes led by MPE (Germany) and including UVIE (Austria);
KUL, CSL, IMEC (Belgium); CEA,  OAMP (France); MPIA (Germany); IFSI, OAP/AOT, OAA/CAISMI, LENS, SISSA
(Italy); IAC (Spain). This development has been supported by the funding agencies BMVIT (Austria),
ESA-PRODEX (Belgium), CEA/CNES (France), DLR (Germany), ASI (Italy), and CICT/MCT (Spain).
This paper is based [in part] on observations made with ESO Telescopes at the Paranal Observatory under 
program ID 179.C-0151. This research has made use of the SIMBAD database, operated at CDS, Strasbourg, France.

\end{acknowledgement}

%++++++++++++++++++++++++++++++++++++++++++++++++++++++++++++++++++++++
{}

%+++++++++++++++++++++++++++++++++++++++++++++++++++++++++++++++++++++++++++++++++
\begin{appendix}

\section{Detection of CO transitions}

\begin{figure*}
   \resizebox{\hsize}{!}{\includegraphics {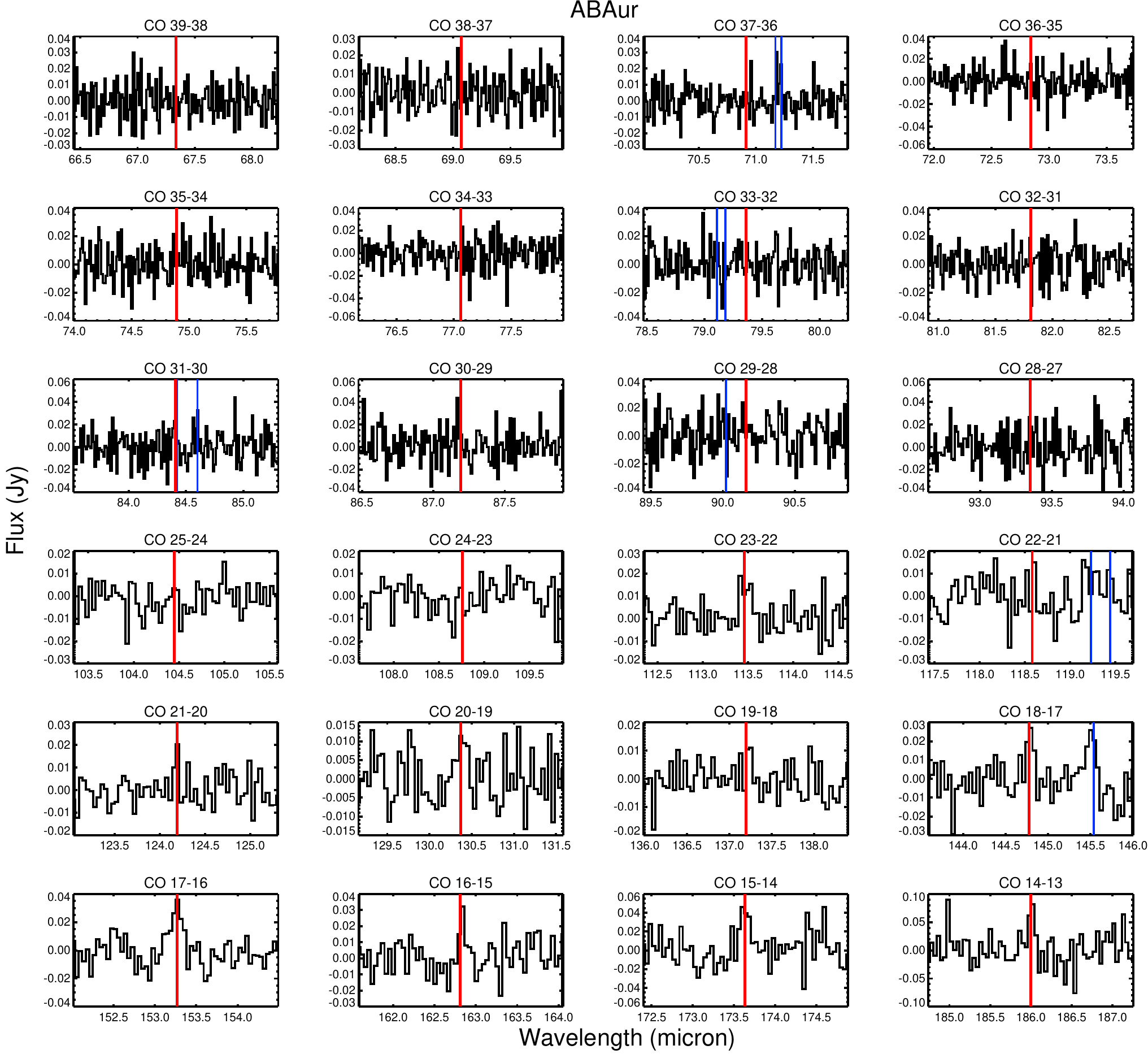}}
\caption{The spectra of AB Aur, centered on the CO lines, indicated in red. The OH line at 84.42 \mic \ blends with the 
CO $J$ = 31 \ra \ 30 line. In blue, we indicate the positions of other lines (CH$^{+}$ at 90 \mic \ and \Oone \ at 145 \mic; rest 
are OH doublets). }
\label{f_spectra1}
\end{figure*}

\begin{figure*}
   \resizebox{\hsize}{!}{\includegraphics {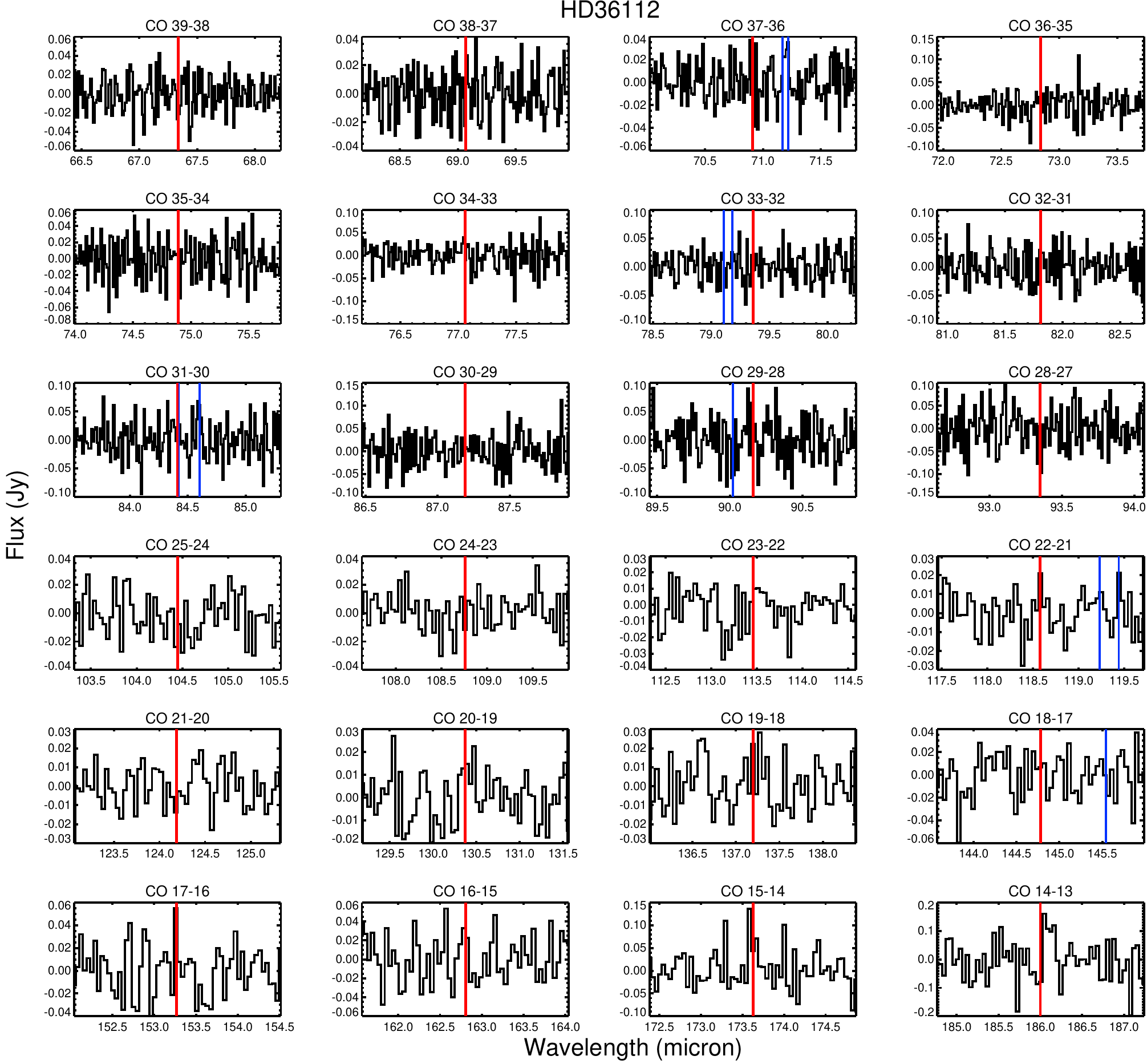}}
\caption{The spectra of HD 36112, centered on the CO lines, indicated in red.The OH line at 84.42 \mic \ blends with the 
CO $J$ = 31 \ra \ 30 line. In blue, we indicate the positions of other lines (CH$^{+}$ at 90 \mic \ and \Oone \ at 145 \mic; rest 
are OH doublets).}
\label{f_spectra2}
\end{figure*}

\begin{figure*}
   \resizebox{\hsize}{!}{\includegraphics {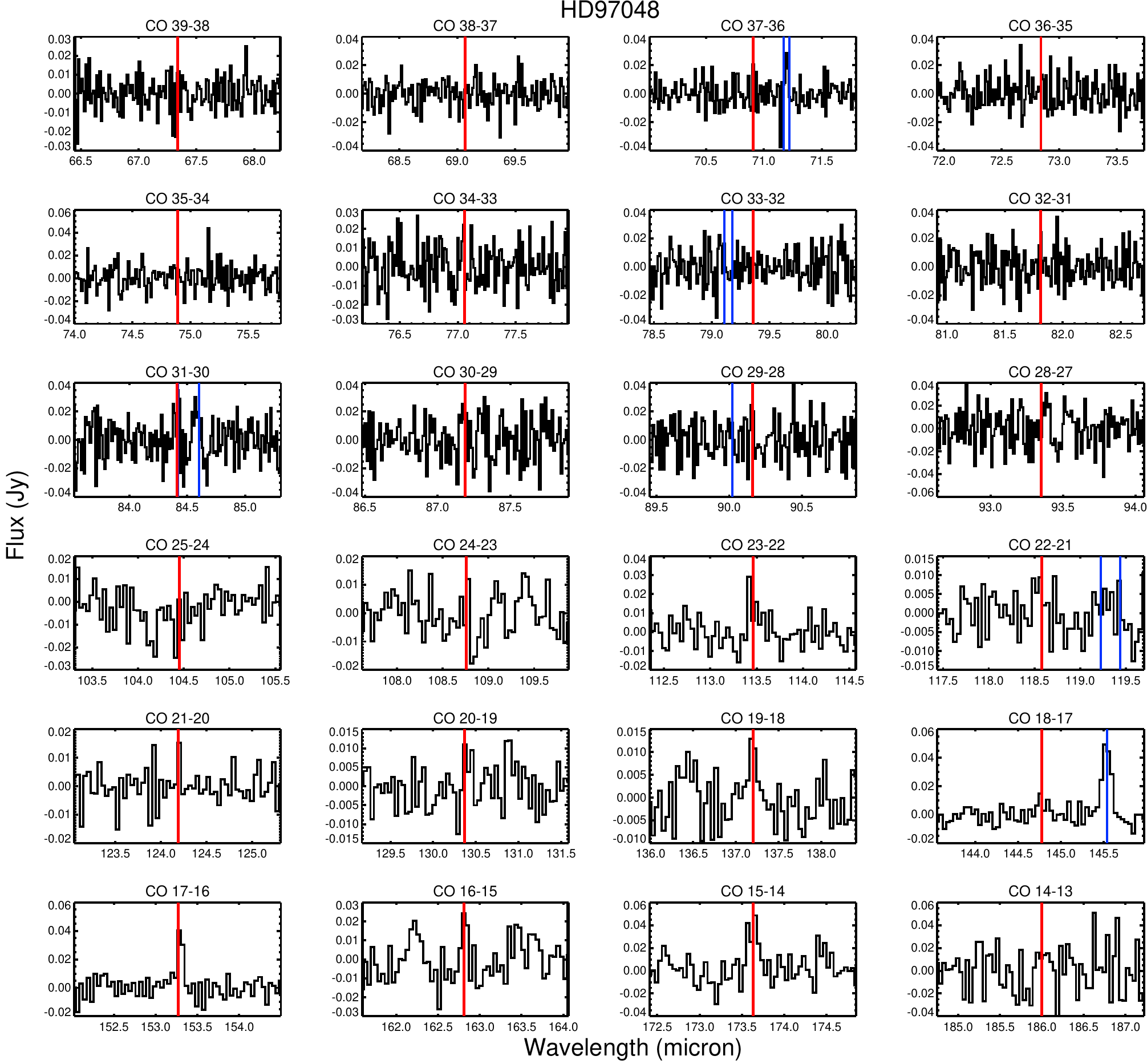}}
\caption{The spectra of HD 97048, centered on the CO lines, indicated in red. The OH line at 84.42 \mic \ blends with the 
CO $J$ = 31 \ra \ 30 line. In blue, we indicate the positions of other lines (CH$^{+}$ at 90 \mic \ and \Oone \ at 145 \mic; rest 
are OH doublets).}
\label{f_spectra3}
\end{figure*}

\begin{figure*}
   \resizebox{\hsize}{!}{\includegraphics {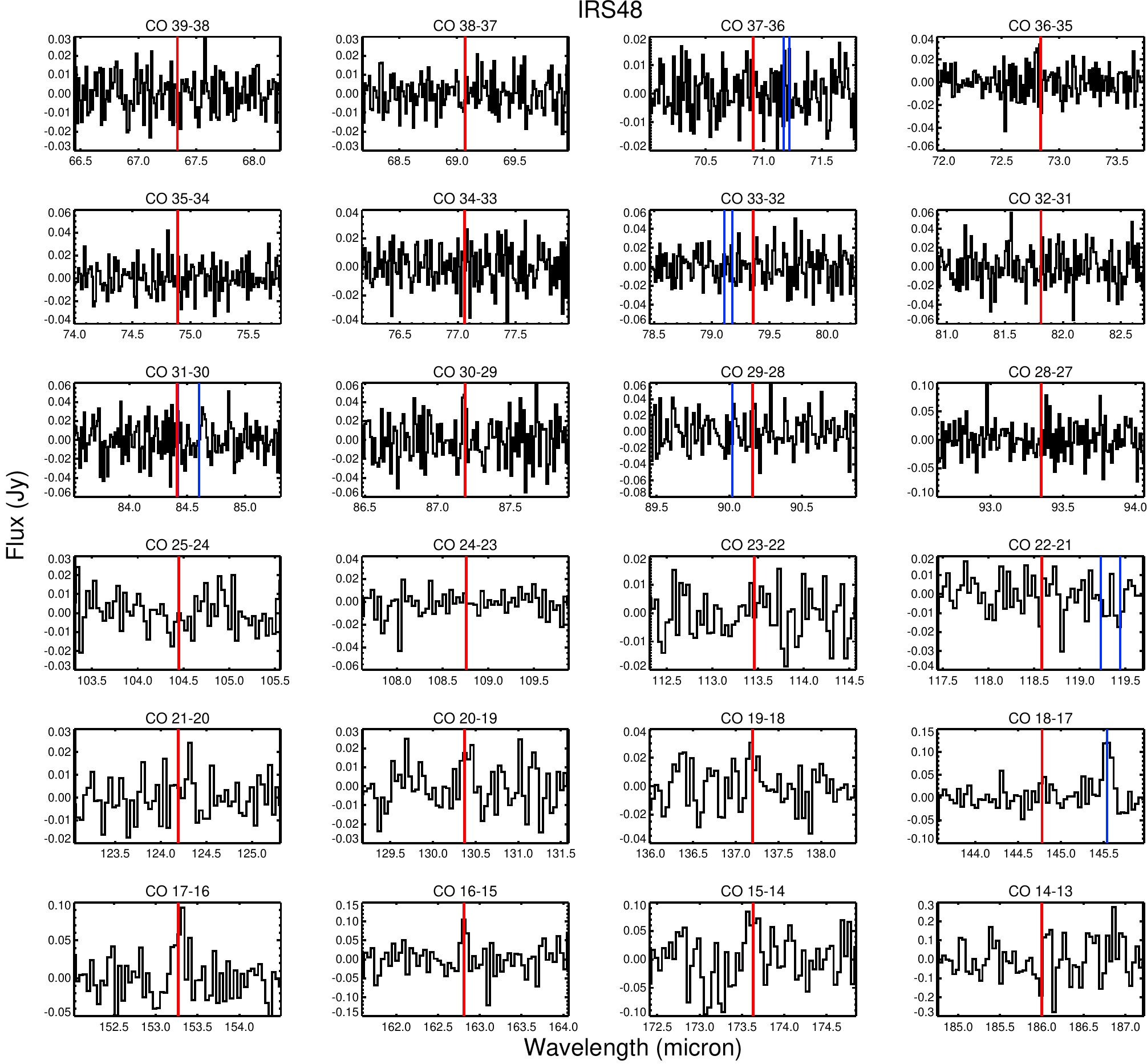}}
\caption{The spectra of IRS 48, centered on the CO lines, indicated in red. The OH line at 84.42 \mic \ blends with the 
CO $J$ = 31 \ra \ 30 line. In blue, we indicate the positions of other lines (CH$^{+}$ at 90 \mic \ and \Oone \ at 145 \mic; rest 
are OH doublets).}
\label{f_spectra4}
\end{figure*}

\begin{figure*}
   \resizebox{\hsize}{!}{\includegraphics {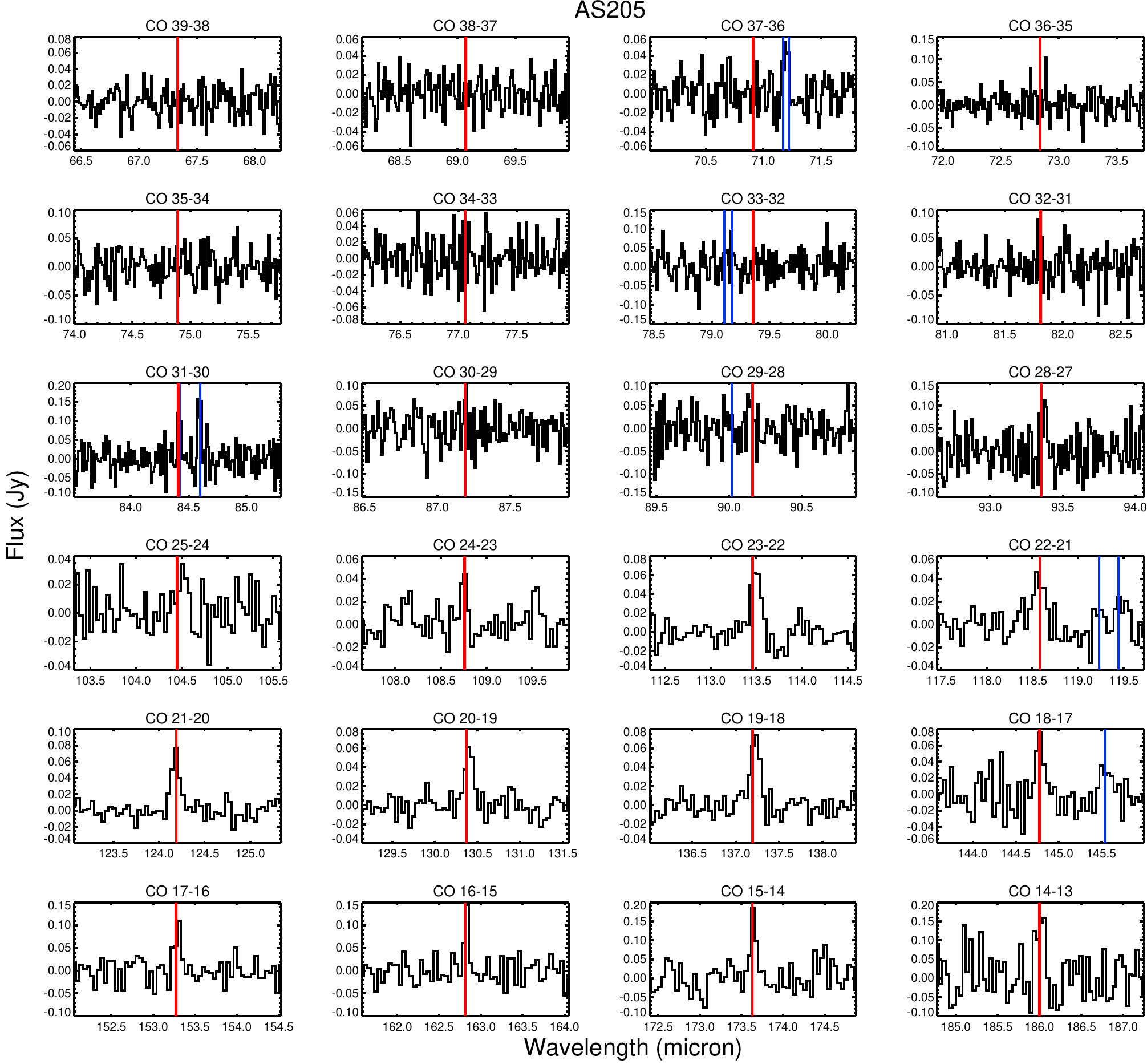}}
\caption{The spectra of AS 205, centered on the CO lines, indicated in red. The OH line at 84.42 \mic \ blends with the 
CO $J$ = 31 \ra \ 30 line. In blue, we indicate the positions of other lines (CH$^{+}$ at 90 \mic \ and \Oone \ at 145 \mic; rest 
are OH doublets).}
\label{f_spectra5}
\end{figure*}

\begin{figure*}
   \resizebox{\hsize}{!}{\includegraphics {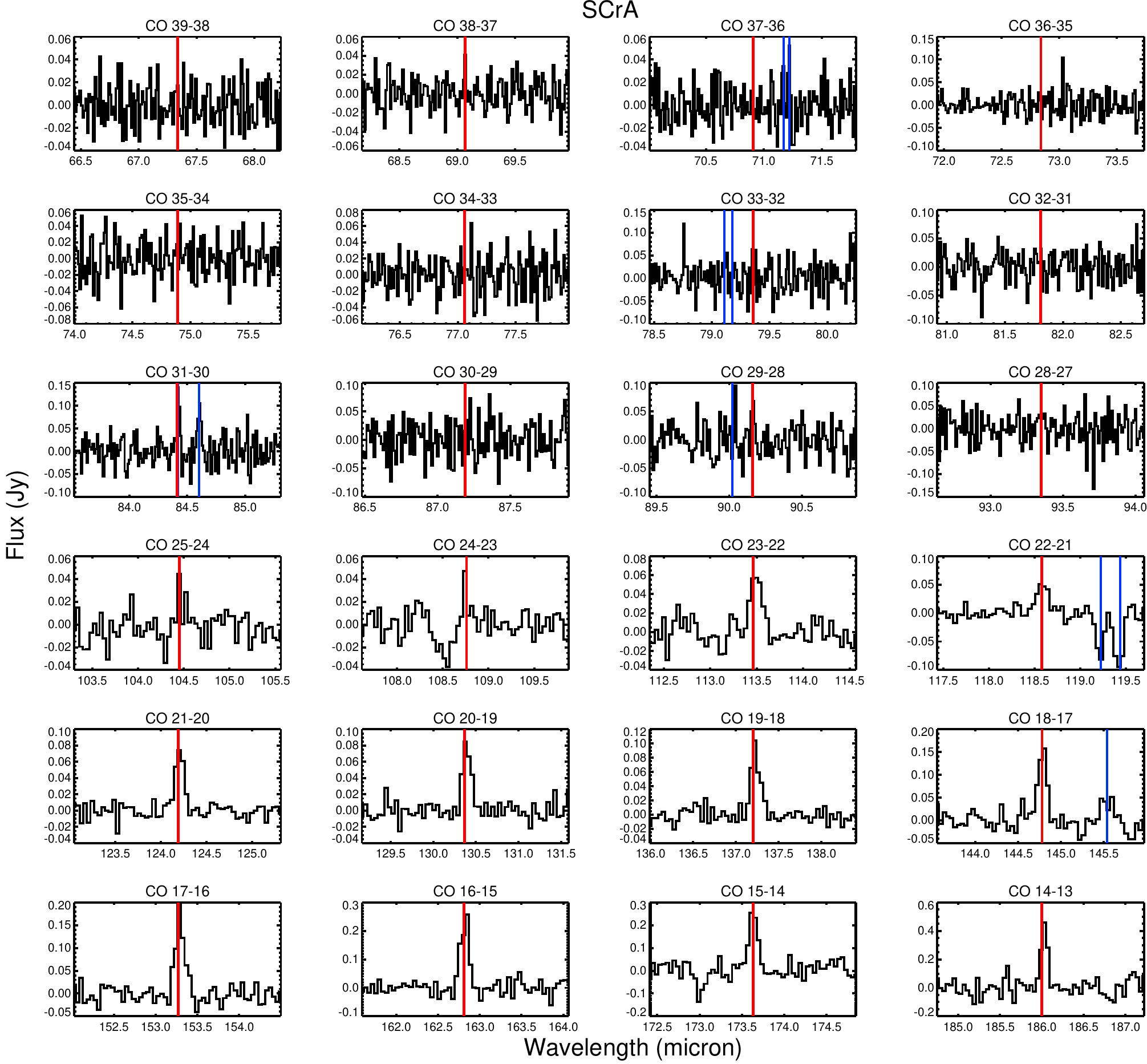}}
\caption{The spectra of S CrA, centered on the CO lines, indicated in red. The OH line at 84.42 \mic \ blends with the 
CO $J$ = 31 \ra \ 30 line. In blue, we indicate the positions of other lines (CH$^{+}$ at 90 \mic \ and \Oone \ at 145 \mic; rest 
are OH doublets).}
\label{f_spectra6}
\end{figure*}

\begin{figure*}
   \resizebox{\hsize}{!}{\includegraphics {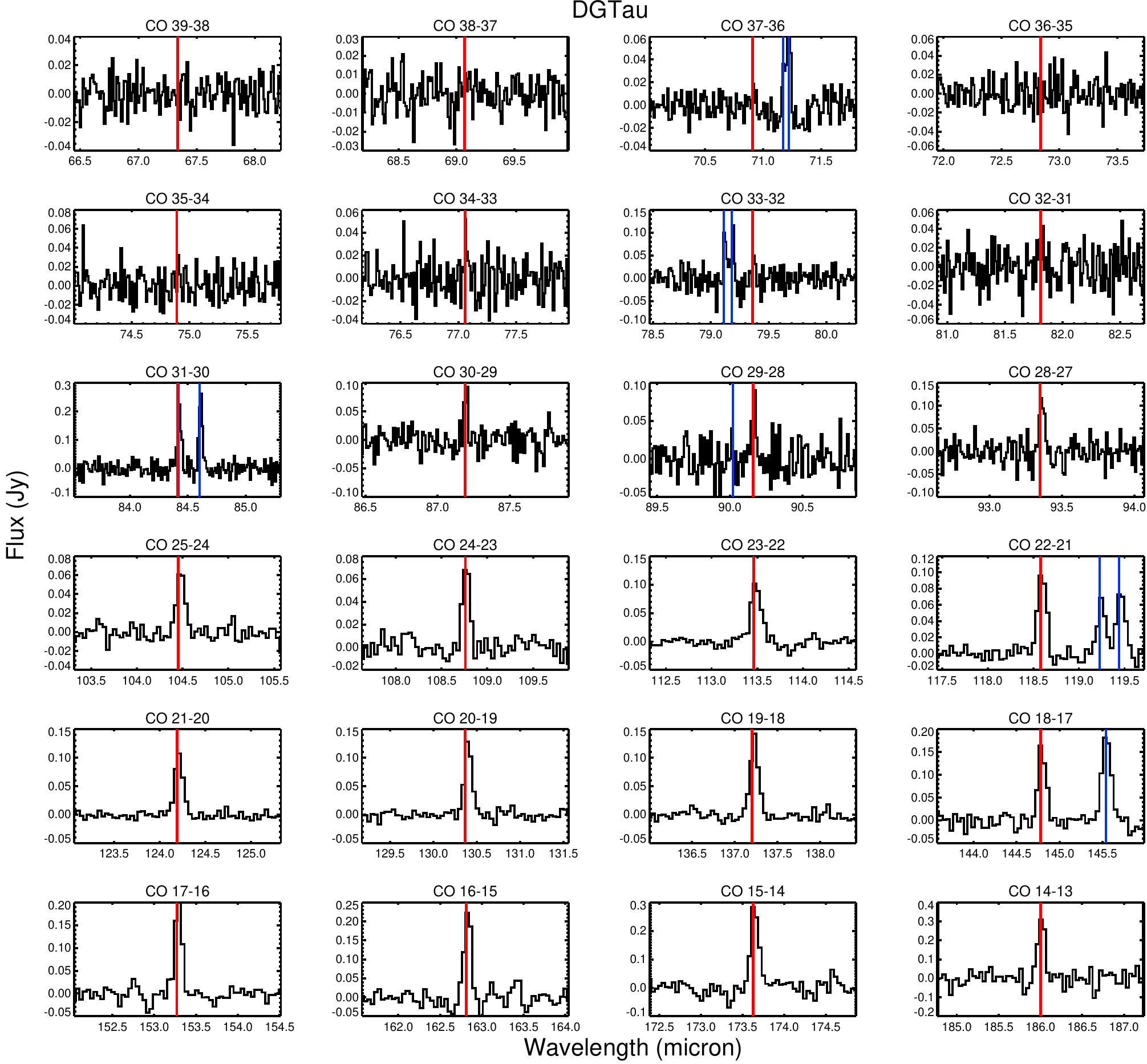}}
\caption{The spectra of DG Tau, centered on the CO lines, indicated in red. The OH line at 84.42 \mic \ blends with the 
CO $J$ = 31 \ra \ 30 line. In blue, we indicate the positions of other lines (CH$^{+}$ at 90 \mic \ and \Oone \ at 145 \mic; rest 
are OH doublets).}
\label{f_spectra7}
\end{figure*}

\begin{figure}[h!]
   \resizebox{\hsize}{!}{\includegraphics {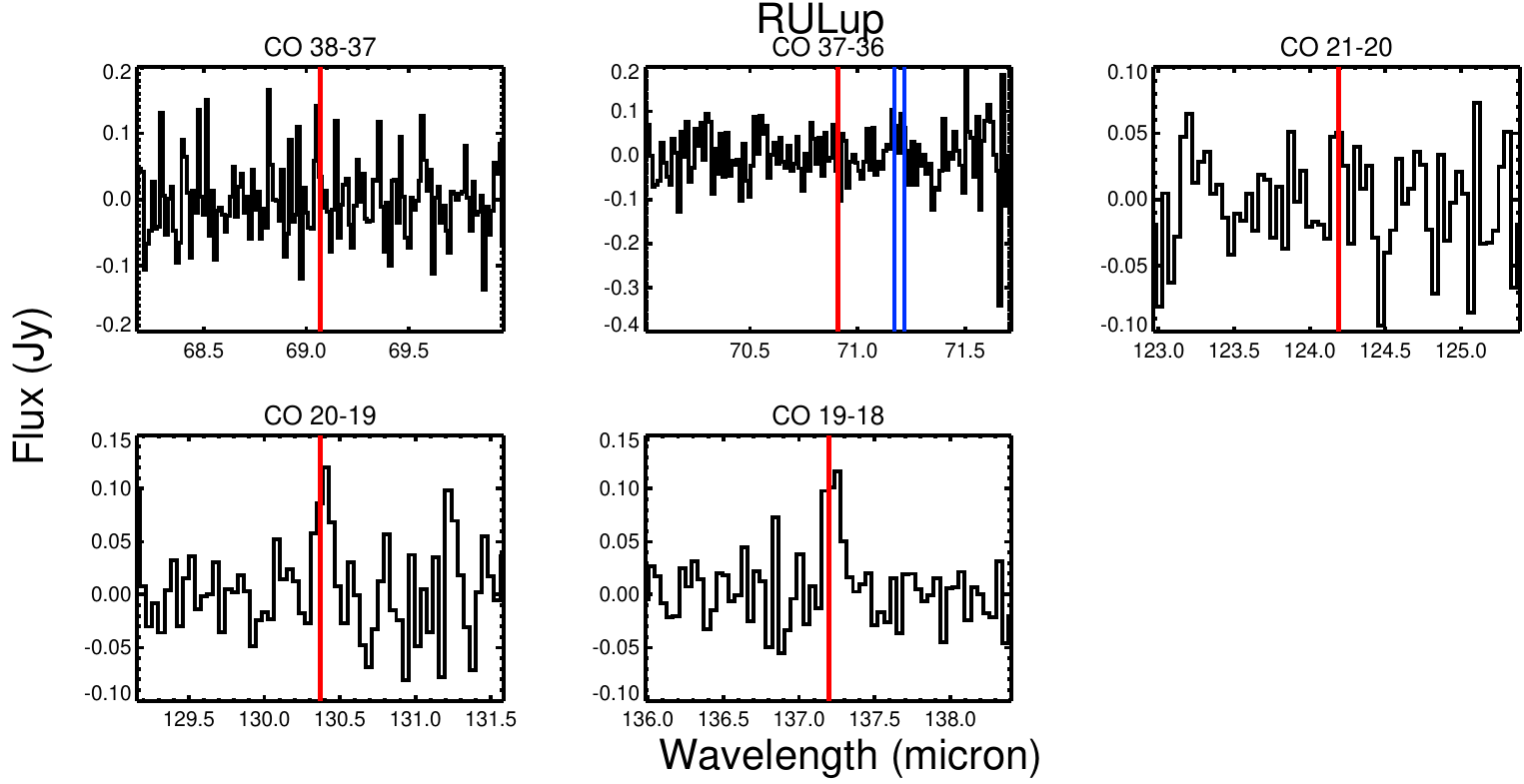}}
\caption{The spectra of RU Lup, centered on the CO lines, indicated in red. In blue, we indicate the position of OH lines.}
\label{f_spectra8}
\end{figure}

In Table~\ref{t_transitions}, we list the transitions that were detected in our PACS spectra, 
while in Figs.~\ref{f_spectra1} to \ref{f_spectra8}, we show windows around the position 
of the CO lines for those objects where CO line emission was detected. 

\begin{table*}
\begin{center}
\caption{CO Transitions detected in the PACS SED scans.}
\begin{tabular}{ccc}
\hline\hline
Transition & Wavelength (\mic) & \eup (K) \\
\hline
$J$36 \ra \ 35  & 72.84 &3669 \\
$J$35 \ra \ 34  & 74.89 &3471 \\
$J$34 \ra \ 33  & 77.11 &3279 \\
$J$33 \ra \ 32  & 79.36 &3093 \\
$J$32 \ra \ 31  & 81.81 &2911 \\
$J$31 \ra \ 30  & 84.41 &2735 \\
$J$30 \ra \ 29  & 87.19 &2565 \\
$J$29 \ra \ 28  & 90.16 &2400 \\
$J$28 \ra \ 27  & 93.35 &2240 \\
$J$25 \ra \ 24  &104.45 &1794 \\
$J$24 \ra \ 23  &108.76 &1657 \\
$J$23 \ra \ 22  &113.46 &1524 \\
$J$22 \ra \ 21  &118.58 &1397 \\
$J$21 \ra \ 20  &124.19 &1276 \\
$J$20 \ra \ 19  &130.37 &1160 \\
$J$19 \ra \ 18  &137.20 &1050 \\
$J$18 \ra \ 17  &144.78 &945  \\
$J$17 \ra \ 16  &153.27 &846  \\
$J$16 \ra \ 15  &162.81 &752  \\
$J$15 \ra \ 14  &173.63 &663  \\
$J$14 \ra \ 13  &186.00 &581  \\
\hline
\end{tabular}
\label{t_transitions}
\end{center}
\end{table*}


\section{Upper limits}

In Tables~\ref{t_uplim}.1 and \ref{t_uplim_forst}.2, we give the upper limits at a few key
transitions ($J$36 \ra \ 35, 29 \ra \ 28, 23 \ra \ 22, 18 \ra \ 17, and 15 \ra \ 14) for those
sources where not a single CO emission line was detected.

\begin{table*}
\begin{center}
\label{t_uplim}
\caption{CO upper limits at selected wavelengths for the sources with full SEDs
scans (50-200 \mic) in W/m$^2 \times 10^{-18}$. The continuum level is given in Jy.}
\begin{tabular}{lcccccccccc}
\hline
Object   &$J$36 \ra \ 35 & Cont.        &$J$29 \ra \ 28 &Cont.         &$J$23\ra \ 22 & Cont.         &$J$18 \ra \ 17 &Cont.          &$J$15 \ra \ 14&Cont.\\
               &                  & 72.8 \mic &                   & 90.2 \mic &                 &113.5 \mic &                   &144.8 \mic &&173.6 \mic \\
\hline
\hline
HD 35187   &$<$ 5.7  & 4.9 &--       &--    &$<$ 16.7 &3.3  &$<$ 10.8 &2.9  &--       &--\\
HD 38120   &$<$ 20.2 & 8.6 &$<$ 25.5 &7.6   &$<$ 11.9 &5.3  &$<$ 9.6  &4.0 &$<$ 13.7  &3.1\\
HD 50138   &$<$ 33.0 &5.7  &$<$ 33.6 &3.9   &$<$ 12.9 &20.1 &$<$ 12.7 &1.22&$<$ 17.1  &0.95\\
HD 100453  &$<$ 67.1 &35.3 &$<$ 20.7 &30.3  &$<$ 14.5 &20.5 &$<$ 9.8  &15.9&$<$ 19.1  &14.0\\
HD 104237  &$<$ 53.9 &9.3  &$<$ 27.6 &8.6   &$<$ 13.9 &6.4  &$<$ 11.7 &4.8&$<$ 19.0  &4.3 \\
HD 135344  &$<$ 24.6 &31.2 &$<$ 19.1 &30.7  &$<$ 12.8 &24.9 &$<$ 6.9  &20.5&$<$ 15.5  &18.2\\
HD 139614  &$<$ 32.8 &18.3 &$<$ 21.7 &17.1  &$<$ 11.2 &13.4 &$<$ 9.5  &12.1&$<$ 17.2  &11.7\\
HD 142527  &$<$ 7.6  &109.7&$<$ 37.2 &95.1  &$<$ 29.6 &74.4 &$<$ 22.1 &63.1&$<$ 37.9  &56.2\\
HD 144668  &$<$ 36.8 &5.6  &$<$ 21.5 &4.5   &$<$ 13.4 &2.8  &$<$ 8.2  &1.9&$<$ 20.1  &1.7\\
HD 163296  &$<$ 4.1  &19.0 &$<$ 23.9 &21.7  &$<$ 13.6 &19.3 &$<$ 11.8 &19.1&16.6 (5.1)&19.4\\
HD 169142  &$<$ 121.3&27.1 &$<$ 53.5 &24.1  &$<$ 33.0 &19.2 &$<$ 20.4 &17.8&$<$ 20.1  &16.1\\
HD 179218  &$<$ 22.8 &21.0 &$<$ 13.9 &16.4  &$<$ 10.1 &10.7 &$<$ 6.5  &7.2&$<$ 12.2  &5.7\\
SR 21    &$<$ 32.3 &33.5 &$<$ 26.4 &29.9  &$<$ 11.0 &21.8 &$<$ 11.3 &16.6 &$<$18.9  &13.9\\
\hline
\end{tabular}
\end{center}
\end{table*}

\begin{table*}
\begin{center}
\label{t_uplim_forst}
\caption{CO Upper limits for the sources with the forsterite scans (60-71 \mic\ + 120-143 \mic) in W/m$^2 \times 10^{-18}$.
With ':' we indicate a detection that is between 2 and 3 sigma. The continuum level is given in Jy.}
\begin{tabular}{lcccc}
\hline
\hline
Object   &CO $J$38 \ra \ 37 & Cont. 70.9 \mic & CO $J$21 \ra \ 20 &Cont. 124\mic \\
\hline
HD 98922   &$<$ 30.8 &6.1    &7.5: (3.2) &1.1\\
HD 141569  &$<$ 22.6 &5.4    &$<$ 13.1   &2.1\\
HD 142666  &$<$ 30.1 &6.7    &$<$ 11.9   &5.1\\
HD 144432  &$<$ 16.7 &5.8    &$<$ 8.0    &2.9\\
HD 150193  &$<$ 24.7 &7.2    &$<$ 7.1    &3.2\\
HT Lup  &$<$ 50.2 &6.2    &$<$ 23.9   &11.0\\
RU Lup  &$<$ 22.0 &5.9    &$<$ 9.0    &4.2   \\
RY Lup  &$<$ 20.3 &6.3    &$<$ 8.6    &3.9\\
RNO90   &$<$ 25.6 &3.8    &$<$ 11.9   &2.1\\
\hline
\end{tabular}
\end{center}
\end{table*}

\end{appendix}


\begin{thebibliography}{}

 \bibitem[2010]{acke2010} Acke, B., Bouwman, J., Juh\'asz, A. et al. 2010, ApJ 718, 558
 
 \bibitem[2008]{agundez2008} Ag\'undez, M., Cernicharo, J. and Goicoechea, J.R. 2008, A\&A 2008, 483, 831

 \bibitem[2008]{bary2008} Bary, J.S., Weintraub, D.A., Shukla, S.J. et al. 2008, ApJ 678, 1088

 \bibitem[2011]{bast2011} Bast, J.E., Brown, J.M., Herczeg G.J. et al. 2011, A\&A 527, 119

 \bibitem[2010]{benisty2010} Benisty, M., Tatulli, E., M\'enard, F. and Swain, M.R. 2010, A\&A 511, A75

 \bibitem[2008]{bitner2008} Bitner, M.A., Richte, M.J., Lacy, J.H. et al. 2008, ApJ 688, 1326

 \bibitem[2004]{blake2004} Blake, G.A. and Boogert, A.C.A. 2004, ApJ 606, 73

 \bibitem[2003]{bouwman2003} Bouwman, J., de Koter, A., Dominik, C. and Waters, L.B.F.M. 2003, A\&A 401, 577

 \bibitem[2007]{brittain2007} Brittain, S.D., Simon, Th. , Najita, J.R., and Rettig, T.W. 2007, ApJ 659, 685

 \bibitem[2012]{brown2012} Brown,  J.M., Herczeg, G.J., Pontoppidan, K.M. and E.F. van Dishoeck 2012, ApJ 744, 116
 
 \bibitem[2013]{brown2013} Brown J. M., Pontoppidan, K. M., van Dishoeck, E.F. et al. 2013, \apj, accepted
 
 \bibitem[2012]{bruderer2012} Bruderer, S., van Dishoeck, E.F., Doty, S.D. and Herczeg, G.J. 2012, A\&A 541, 91

 \bibitem[2008]{carmona2008} Carmona, A., van den Ancker, M.E., Henning, Th. et al. 2008, A\&A 477, 839

 \bibitem[2011]{carmona2011} Carmona, A., van der Plas, G., van den Ancker, M.E. et al. 2011, A\&A 533, A39

 \bibitem[2005]{dent2005} Dent, W.R.F., Greaves, J.S., Coulson, I.M. 2005, MNRAS 359, 663

 \bibitem[2003]{dominik2003} Dominik,\,C., Dullemond,\,C.P., Waters,\,L.B.F.M. et al. 2003, A\&A 398, 607

 \bibitem[2001]{dullemond2001} Dullemond,\,C.P., Dominik,\,C. and Natta, A. 2001, \apj, 560, 957

 \bibitem[2012]{france2012} France, K., Schindhelm, E., Herczeg, G.~J., et al.\ 2012, \apj, 756, 171 

 \bibitem[2006]{geers2006} Geers, V.C., Augereau, J.-C.,  Pontoppidan, K.M. et al. 2006, A\&A 459, 545
 
 \bibitem[2007]{geers2007} Geers, V.C., Pontoppidan, K.M., van Dishoeck, E.F. et al. 2007, A\&A 476, 279

 \bibitem[1999]{giannini1999} Giannini, T., Lorenzetti, D., Tommasi, E. et al. A\&A 346, 617

 \bibitem[1999]{goldsmith1999} Goldsmith, P.F., and Langer, W.D. 1999, ApJ 517, 209

 \bibitem[2006]{goto2006} Goto, M., Usuda, T., Dullemond, C.P. et al. 2006, ApJ 652, 758

 \bibitem[2013]{green2013} Green, J. D., Evans, N. J. II, J\"orgensen, J. K. et al. 2013, \apj, accepted (arXiv:1304.7389)

 \bibitem[2009]{hersant2009} Hersant, F., Wakelam, V., Dutrey, A. et al. 2009, A\&A 493, L49

 \bibitem[2005]{herczeg2005} Herczeg, G. J., Walter, F.M., Linsky, J.L. et al. 2005, AJ 129, 2777

 \bibitem[2006]{herczeg2006} Herczeg, G. J., Linsky, J.L., Jeffrey, L. et al. 2006, ApJS 165, 256

 \bibitem[2010]{honda2010} Honda, M., Inoue, A.K., Okamoto, Y.K. et al. 2010, \apj 718, 199

 \bibitem[2011]{ingleby2011} Ingelby, L., Calvet, N., Bergin, E. et al. 2011, ApJ 743, 105

 \bibitem[2010]{isella2010} Isella, A., Natta, A., Wilner, D., Carpenter, J.M. and Testi, L. 2010, \apj 725, 1735
 
 \bibitem[2007]{jonkheid2007} Jonkheid, B., Dullemond, C.~P., Hogerheijde M.~R., \& van Dishoeck,
  E.~F. 2007, \aap, 463, 203

 \bibitem[2010]{juhasz2010} Juh\'asz, A., Bouwman, J. Henning, Th. et al. 2010, ApJ 721, 431
 
 \bibitem[2013]{karska2013} Karska, A., Herczeg, G.J., van Dishoeck, E.F., et al. 2013, A\&A 552, 141
 
 \bibitem[2004]{kamp2004} Kamp, I., and Dullemond, C.P. 2004, \apj 615, 991
 
 \bibitem[2011]{kamp2011} Kamp, I, Woitke, P., Pinte, C. et al. 2011, A\&A 532, 85 

 \bibitem[1995]{koerner1995} Koerner, D.W., and Sargent, A.I. 1995, AJ 109, 2138

 \bibitem[2007]{lahuis2007} Lahuis,  F., van Dishoeck, E.F., Blake, G.A. et al. 2007, ApJ 665, 492
 
 \bibitem[2000]{lavalley2000} Lavalley-Fouquet, C., Cabrit, S. and Dougados, C. 2000, A\&A 356, L41
 
 \bibitem[2013]{maaskant2013} Maaskant, K. M., Honda, M., Waters, L.B.F.M. et al. 2013, arXiv:1305.3138

 \bibitem[2009]{claire2009} Martin-Za\"idi, C., Habart, E., Augereau, J.-C. et al. 2009, ApJ, 695, 1302

 \bibitem[2010]{claire2010} Martin-Za\"idi, C., Augereau, J.-C., M\'enard, F. et al. 2010, A\&A 516, A110

 \bibitem[2001]{meeus2001} Meeus, G., Waters, L.B.F.M., Bouwman, J. et al. 2001, A\&A 365, 476

 \bibitem[2012]{meeus2012} Meeus, G., Montesinos, B., Mendigut\'ia, I. et al. 2012, A\&A 544, 78

 \bibitem[1984]{mundt1984} Mundt R. 1984, \apj 280, 749

 \bibitem[2003]{najita2003} Najita, J., Carr, J.S., and Mathieu, R.D. 2003, ApJ 589, 931

 \bibitem[2010]{oberg2010} \"Oberg, K.I., Qi, C., Fogel, J.K.J. et al. 2010, ApJ 720, 480

 \bibitem[2008]{panic2008} Pani\'c, O., Hogerheijde, M.R., Wilner, D., and Qi, C. 2008, A\&A 491, 219

 \bibitem[2009]{panic2009} Pani\'c, O. and Hogerheijde 2009, A\&A 508, 707

 \bibitem[2010]{panic2010} Pani{\'c}, O., van Dishoeck, E.~F., Hogerheijde, M.~R., et~al. 2010,  \aap, 519, A110

 \bibitem[2010]{pilbratt2010} Pilbratt, G. et al. 2010, A\&A 518, L1

 \bibitem[2012]{podio2012} Podio, L., Kamp, I. et al. 2012, A\&A, 545, 44

 \bibitem[2010]{poglitsch2010} Poglitsch et al. 2010, A\&A 518, L2

 \bibitem[2008]{pontoppidan2008} Pontoppidan K.M., Blake, G.A., van Dishoeck, E.F. et al. 2008, ApJ 684, 1323
 
 \bibitem[2011]{pontoppidan2011} Pontoppidan K.M., van Dishoeck, E.F., Blake, G.A. et al. 2011, Msngr, 143, 32

 \bibitem[2007]{rollig2007} R{\"o}llig, M., Abel, N.~P., Bell, T., et~al. 2007, \aap, 467, 187

 \bibitem[2009]{salyk2009} Salyk, C., Blake, G.A., Boogert, A.C.A. and Brown, J.M. 2009, ApJ 699, 330

 \bibitem[2011]{salyk2011} Salyk, C., Blake, G.A., Boogert, A.C.A., and Brown, J.M. 2011, ApJ 743, 112 

 \bibitem[2013]{salyk2013} Salyk, C., Herczeg, G.J., Brown, J.M. et al. 2013, \apj 769, 21

 \bibitem[2010]{sturm2010} Sturm, B., Bouwman, J., Henning, T., et al.\  2010, \aap, 518, L129 

 \bibitem[2001]{thi2001} Thi, W.~F., van Dishoeck, E.~F., Blake, G.~A., et al. 2001, ApJ, 561, 1074

 \bibitem[2011]{thi2011} Thi, W.~F., M\'enard, F., Meeus, G. et al. 2011, A\&A 530, L2
 
 \bibitem[2013]{thi2013} Thi, W.~F., Kamp, I., Woitke, P. et al. 2013, A\&A 551, 49
 
 \bibitem[2013]{marel2013} van de Marel, N., van Dishoeck, E.F., Bruderer, S. et al. 2013, Sci 340, 1199

 \bibitem[2009]{plas2009} van der Plas, G., van den Ancker, M.E., Acke, B. et al. 2009, A\&A 500, 1137

 \bibitem[2001]{weingartner2001} Weingartner, J.~C. \& Draine, B.~T. 2001, \apj, 548, 296
 
 \bibitem[2011]{williams2011} Williams, J.P. and Cieza, L.A. 2011, ARA\&A 49, 67

 \bibitem[2010]{woitke2010} Woitke, P., Pinte, C., Tilling, I., et~al. 2010, \mnras \ 405, L26

 \bibitem[2012]{yang2012} Yang, H., Herczeg, G.J., Linsky, J.L. et al. 2012, ApJ 744, 121

\end{thebibliography}
\end{document}